\documentclass[]{aa}

\usepackage{graphicx}
\usepackage{txfonts}
\usepackage{natbib}
\usepackage{hyperref}
\usepackage{tikz}
\usetikzlibrary{fit, backgrounds}
\usetikzlibrary{shapes,fit}
\usepackage{xcolor}
\usepackage{verbatim}
\usepackage{listings}
\usepackage{hyperref}
\usepackage{cleveref}
\usepackage{mathrsfs}
\usepackage{dsfont}
\usepackage{amsmath}
\usepackage{graphicx}
\usepackage{subcaption}
\usepackage{enumitem}

\AtBeginShipout{%
	\ifnum\value{page}>1 %
	\typeout{* Additional boxing of page `\thepage'}%
	\setbox\AtBeginShipoutBox=\hbox{\copy\AtBeginShipoutBox}%
	\fi
}

\bibpunct{(}{)}{;}{a}{}{,} 

\def\aap{Astronomy and Astrophysics} 
\def\apjl{Astrophysical Journal Letters}
\def\apjs{The Astrophysical Journal Supplement}

\renewcommand{\vec}[1]{\boldsymbol{#1}}

\DeclareMathOperator{\Tr}{\text{Tr}}
\DeclareMathOperator{\diag}{\text{diag}}

\begin{document}
 \title{Denoising, deconvolving and decomposing multi-domain photon observations}
 \titlerunning{D$^4$PO}
   \subtitle{The D$^4$PO algorithm}
   \author{Daniel Pumpe
   	       \and Martin Reinecke
                \and Torsten A. En{\ss}lin}
	\authorrunning{Daniel Pumpe et. al. }
   \institute{Max Planck Institute for Astrophysics, 
              Karl-Schwarzschild-Str. 1, 
              D-85741 Garching, Germany
             }

   \date{Received 06 February, 2018; accepted 23 August, 2018}
 
  \abstract   {Astronomical imaging based on photon count data is a non-trivial task. In this context we show how to denoise, deconvolve, and decompose multi-domain photon observations. The primary objective is to incorporate accurate and well motivated likelihood and prior models in order to give reliable estimates about morphologically different but superimposed photon flux components present in the data set. Thereby we denoise and deconvolve photon counts, while simultaneously decomposing them into diffuse, point-like and uninteresting background radiation fluxes. The decomposition is based on a probabilistic hierarchical Bayesian parameter model within the framework of information field theory (IFT). In contrast to its predecessor D$^3$PO, D$^4$PO reconstructs multi-domain components. Thereby each component is defined over its own direct product of multiple independent domains, for example location and energy. D$^4$PO has the capability to reconstruct correlation structures over each of the sub-domains of a component separately. Thereby the inferred correlations implicitly define the morphologically different source components, except for the spatial correlations of the point-like flux. Point-like source fluxes are spatially uncorrelated by definition. The capabilities of the algorithm are demonstrated by means of a synthetic, but realistic, mock data set, providing spectral and spatial information about each detected photon. D$^4$PO successfully denoised, deconvolved, and decomposed a photon count image into diffuse, point-like and background flux, each being functions of location as well as energy. Moreover, uncertainty estimates of the reconstructed fields as well as of their correlation structure are provided employing their posterior density function and accounting for the manifolds the domains reside on.}


   \maketitle

\section{Introduction}

In most astrophysical data sets arising from photon counting observations, 
different physical fluxes superimpose each other. Point-like sources,
compact objects, diffuse emission, background radiation, and other sky structures 
imprint on the data. Furthermore multiple
instrumental distortions such as an imperfect instrument response and noise
complicate the data interpretation. In particular the data of high energy photon
and particle telescopes are subject to Poissonian shot noise which turns any
smooth emission region into a granular image in the detector plane. In this paper, 
a novel technique is developed with the ultimate goal to provide reliable estimates of the actual sky photon flux components, diffuse and point-like, which both vary as a function of location and photon energy. This is achieved by a rigorous 
mathematical and statistical treatment. \\

Individual photon counts are subject to Poissonian shot noise, whose
amplitude depends on the count rate itself. The signal-to-noise ratio (S/N)
drops for low count rates, limiting the detection and
discrimination of faint sources. The ill-posed inverse problem is complicated also by the fact 
that telescopes are inexact in the sense that the exposure is inhomogeneous
across their field of view. In addition, the instrument response function may be
non-linear and only known up to a certain accuracy.
Especially point sources are smeared out in the data plane by the
instrument's point spread function (PSF), which might let the point sources appear as extended
objects in the data plane. \\

The observed photon flux
is a superposition of multiple morphologically different emission structures
that are best decomposed into their original components to better understand
their causes. Typical morphologically different emitters in astrophysics can be
characterised as diffuse, compact, and point-like sources. On the one hand, diffuse objects illuminate the sky over extended areas and show distinct spatial and spectrally extended structures, for example galaxy clusters. On the other hand, point sources, as main sequence stars, are local features that do not show any spatial structure by definition. They typically have well structured broad-band
energy spectra. Compact objects, for example cool core clusters in between the extremes of point-like
sources and diffuse emission regions, will not be considered here as a separate
morphological class. They should either be regarded as part of the diffuse or
the point-like flux. In addition to the detected cosmic sources, 
the recorded counts might contain events due to background
radiation (such as instrumental background, charged particle induced background,
scattered solar radiation, cosmic rays and other unwanted signals). If its morphological
structure is significantly different from that caused by diffuse and/or point-like
sources, one may be able to distinguish it from sky emission. Overall, this
leads to the obvious question how to denoise, deconvolve, and decompose the
observed data set into its original emission components. The ill-posed inverse problem arises because 
no unique solution exists to split the observed counts into the three accounted components, 
i.e. diffuse, point-like source and background radiation fluxes. 
To tackle this generic inference problem in astrophysics, we introduce D$^4$PO to \textbf{d}enoise, \textbf{d}econcolve, and \textbf{d}ecompose multi-\textbf{d}omain photon observations. In this context we aim to reconstruct the outlined flux components not only in the spatial domain, but also simultaneously with respect to other domains like energy or time. This requires that spatial and spectral or temporal information is available for each detected photon or at least most of them. We show how to infer multiple morphologically different photon flux components while each component can depend on multiple domains, such as energy, time and location. 
As the decomposition of superimposed signals is degenerate, a detector event can in principle be assigned to any of the considered components. Therefore additional information has to be folded into the inference. For this, D$^4$PO relies on a Bayesian approach which allows in a natural way to incorporate a valid data model and a priori knowledge about the
source structures to break the degeneracy between the different flux
contributions. \\ \\ 

First attempts in this direction have been pursued by a maximum likelihood
analysis \citep{article}, followed by maximum entropy analysis
\citep{1985A&A...143...77C,2003A&amp;A...411L.127S} and least\-square techniques for sparse systems
\citep{2013A&amp;C.....1...59B} which were applied to various astrophysical
photon count data sets, such as INTEGRAL/SPI \citep{2003A&amp;A...411L..63V},
COMPTEL \citep{1993ApJS...86..657S} etc..
A popular tool to separate point-like sources from diffuse emission regions is
SExtractor \citep{1996A&amp;AS..117..393B}, which provides a catalogue of point-like
sources. However SExtractor is not highly sensitive in the detection of faint and extended sources \citep{2018arXiv180405591K}, but it was successfully applied in the X-ray regime on filtered images \citep{Valtchanov_2001}. To identify point-like sources in ROSAT X-ray photon observations \citep{1999A&A...349..389V}, which includes background radiation, \cite{2001A&A...370..649B} developed an inference technique based on a maximum likelihood estimator for each detected photon, taking into account energy and position. 
Thereby they assume a parametric Poisson model for the likelihood and can therefore denoise and identify point-like sources in the data set. 
Within the field of sparse regularisation
multiple techniques exploiting an assumed signal sparsity with respect to various functional basis systems or waveforms have been proven to
successfully denoise and deconvolve different types of data
\citep{2006MNRAS.369.1603G,2007ASPC..371..247W,2009ITIP...18..310D,2010ITIP...19.1720B,willett,2017OptLT..90..128L,2016FNL....1550007O,2014arXiv1412.2210V,2014SPIE.9019E..0BD,2010ITIP...19.3133F}.

Disregarding the Poissonian statistics of photon counts, a generic method to
denoise, deconvolve and decompose simulated radio data assuming Gaussian noise
statistics has been developed \citep{2008StMet...5..307B,2013MNRAS.429..165C}.
Further in the regime of Gaussian noise statistics,
\cite{2005A&amp;A...439..401G} developed an algorithm to decompose point and extended sources based on the minimisation of least squares residuals. The CLEAN algorithm \citep{1974A&amp;AS...15..417H},
widely used in radio interferometery, assumes that the whole sky emission is composed of
point-like sources, which leads to inferior reconstructions of the diffuse and
background emission \citep{2007astro.ph..1171S}. Extensions of CLEAN to model the sky with Gaussian blobs to
account for diffuse emission structures have improved on this
\citep{2009GGG....10.9U07A,2011A&amp;A...532A..71R}, however it is still unclear how to optimally set the width of these Gaussians.
The algorithm PowellSnakes I/II \citep{2009MNRAS.393..681C,2012MNRAS.427.1384C} was successfully applied on the Planck sky data \citep{2011A&amp;A...536A...7P}. It is capable of analysing multi-frequency data and to detect point-like sources within diffuse emission regions.  \\
A Bayesian approach close to ours has been developed to separate the background signal from the sky sources in the X-ray part of the electromagnetic spectrum \citep{2009MNRAS.396..165G, 2004ASPC..314..253G}. This method is based on a two-component mixture model which jointly infers point-like sources and diffuse emissions including their uncertainties. \\

This work builds on the D$^3$PO algorithm \citep{Selig:2015rt},
which was successfully applied to the FERMI LAT data \citep{Selig:2015ul}. D$^3$PO denoises,
deconvolves and decomposes photon counts into two signals, a point-like and a
diffuse one, while it simultaneously reconstructs the spatial power spectrum of
the latter. This is done through a hierarchical parameter model incorporating
prior knowledge. \\ \\

D$^4$PO advances upon D$^3$PO in several ways. First, the number of components to be inferred is extended to be more than two. Second, the components can live over different domains: a brightness distribution on the sky can be inferred jointly with a background level living only in a temporal domain. Third, and most importantly, the manifolds, i.e. where the different components live on, can be product manifolds from different domains, as a flux can be a function of location, time, and energy. 

Following the work of \citealt{Selig:2015rt}, we derive the algorithm within the framework of information field
theory (IFT, \citealt{2009PhRvD..80j5005E}), which allows in a natural way the
incorporation of priors. This prior
knowledge is crucial as it is used to discriminate between the morphologically
different sources via their individual statistical properties. While D$^3$PO
could reconstruct the diffuse component depending purely on its location, we
show how to incorporate further information present in the data to get reconstructions that do not
only depend on the location of the reconstruction but also on its energy and/or time. 

All fluxes, be they diffuse, point-like or background, are modelled individually
as signal fields. A field is a continuous quantity defined over a continuous
space. A space here is the domain of one or several manifolds or sub-domains.
For example the sky emissivity is regarded to be a field living over the product
of the two dimensional angular manifold of the celestial sphere times a
one-dimensional spectral energy manifold. Diffuse continuum emission is a smooth
function of both sky position and photon energy. Within each sub-manifold of a
field space we assume statistical homogeneity and isotropy separately for the
field. This joint field correlation over the composed space is assumed to be a
direct product of the sub-domain correlations. This provides the novel
possibility to reconstruct fields which do not only depend on one parameter (in
this case, location), but on multiple parameters, such as location and energy.

\begin{figure*}[]
\centering
\includegraphics[width=\textwidth]{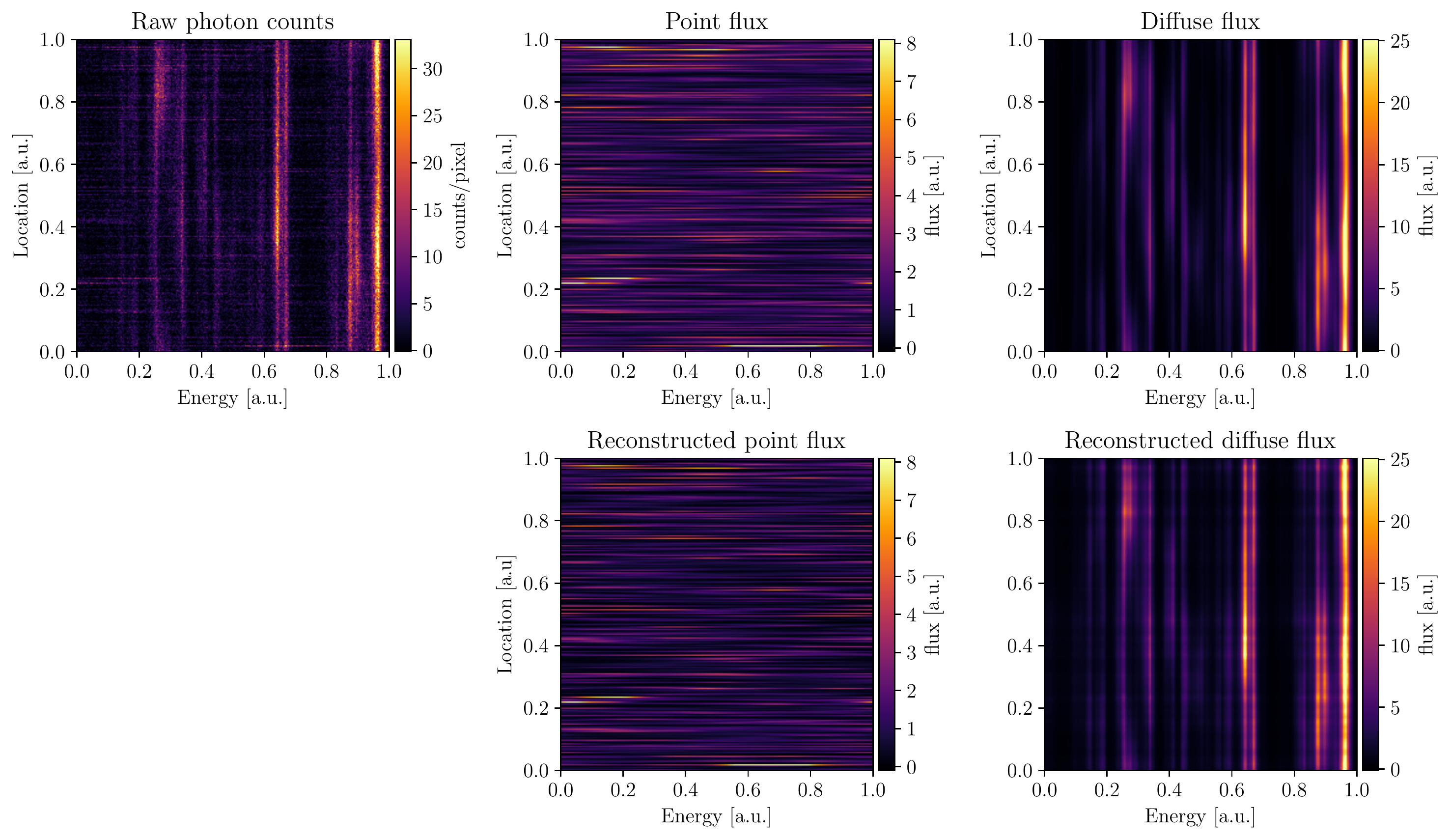}
\caption{Showcase for the D$^4$PO algorithm to denoise, deconvolve, and decompose multi-domain photon observations. This is a  simplified test scenario, which disregards a potential background flux and has only a unit response function (a perfect PSF). The top panel shows (from left to right) the data, providing spatial versus spectral information about each detected photon, the original multi-domain point-like flux, as well as the multi-domain diffuse fields. The bottom panels illustrate the reconstruction of both fields. All fields are living over a regular grid with $350 \times 350$ pixels. }
\label{fig:teaser_su}
\end{figure*} 
Figure~\ref{fig:teaser_su} illustrates such a reconstruction scenario, the
denoising, deconvolving, and decomposition of an exemplary data set generated by
an emission that consists of diffuse and point-like fluxes, each living over
two different sub-domains. Thereby the data set contains information about photon observations including their location and energy. The first sub-domain of both photon flux components is a one-dimensional continuous space representing the spatial domain, while the second sub-domain represents the spectral information of the photons. The exact statistical properties of both components will be discussed in detail in Sec.~\ref{subsec:prior}, but from Fig.~\ref{fig:teaser_su} our implicit understanding of diffuse and point-like fluxes, which depend on energy and location, becomes apparent. 
Here the spatial correlation length of the diffuse flux in the spatial
sub-domain is significantly longer compared to that of the
point-like sources, which only illuminate the sky at distinct locations. In contrast to this
are the correlations in the energy sub domain, where the point-like sources shine over
a broader energy range compared to the diffuse emission.\\
The numerical implementation of the algorithm is done in \textsc{NIFTy3}
\citep{Steininger:2017hb,2013A&amp;A...554A..26S}. This allows us to set up the
algorithm in a rather abstract way by being independent of a concrete data set, domain
discretisation and number of fields. Thanks to the abstractness and the
flexibility of \textsc{NIFTy3}, reconstructions such as those shown in
Fig.~\ref{fig:teaser_su} can easily be extended to different domains (i.e. the
sphere) and to more source components, i.e. additional backgrounds. \\

The structure of this work is as follows: In Sec.~\ref{sec:inference} we give a
detailed derivation of the outlined algorithm, with an in-depth discussion of the
incorporated models and priors. Section~\ref{sec:Inference} describes different
approaches to calculate the derived posterior probability density function (PDF). In Sec.~\ref{sec:Algorithm} we
discuss numerically efficient implementations of the algorithm. Its performance
is demonstrated in Sec.~\ref{sec:mock_example} by its application to a simulated astrophysical data set, whose results are compared to a Maximum Likelihood approach. Finally we conclude in Sec.~\ref{sec:conclusion}.

\section{Inference from photon observation}
\label{sec:inference}
\subsection{Signal inference}
Our goal is to image the high energy sky based on photon count data as it is provided by the astroparticle physics instruments like Fermi \citep{2009ApJ...697.1071A}, Integral \citep{2003A&amp;A...411L..63V}, Comptel \citep{1993ApJS...86..657S}, CTA \citep{2013APh....43....3A}. The sky emissivity is a function of the sky position and photon energy, that is generated by a number of spectral and morphologically different sources.\\
Due to experimental constraints and practical limitations (such as limited observation time, energy range, and spatial and spectral resolution), the obtained data set of any photon count experiment cannot capture all degrees of freedom of the underlying photon flux. As a physical photon flux is a continuous scalar field that can vary with respect to various parameters, such as time, location and energy, we let all signals of interest live in a continuous space over some domain $\Omega$.
Hence we are facing an underdetermined inference problem as there are infinitely many signal field configurations leading to the same finite data set. Consequently we need to use probabilistic data analysis methods, which do not necessarily provide the physically correct underlying signal field configuration but provide expectations and remaining uncertainties of the signal field. 

In this context we are investigating the a posteriori probability distribution $P(\vec{\varphi}\vert \vec{d})$, which states how likely a potential signal $\vec{\varphi}$ is given the data set $\vec{d}$. This is provided by Bayes' theorem
\begin{equation}
P (\vec{\varphi} \vert \vec{d}) = \frac{P(\vec{d} \vert \vec{\varphi}) P(\vec{\varphi})}{P(\vec{d})}\,,
\label{eq:Bayes}
\end{equation}
which is the quotient of the product of the likelihood PDF $P(\vec{d} \vert \vec{\varphi})$ and the signal prior PDF $P(\vec{\varphi})$ divided by the evidence PDF $P(\vec{d})$. The likelihood describes how likely it was to observe the measured data set $\vec{d}$ given a signal field $\vec{\varphi}$. It covers all processes that are relevant for the measurement. The prior describes all a priori knowledge on $\vec{\varphi}$ and must therefore not depend on $\vec{d}$ itself. 
We estimate the posteriori mean $\vec{m}$ of the signal field given the data and its uncertainty covariance $\vec{D}$, which are defined as
\begin{align}
\vec{m} &= \langle \vec{\varphi}\rangle_{(\vec{\varphi} \vert{d})} = \int \mathcal{D}\vec{\varphi}\, \vec{\varphi} P (\vec{\varphi}\vert \vec{d})\, , \quad \text{and}
\label{eq:m} \\
\vec{D} &= \langle (\vec{m}-\vec{\varphi})(\vec{m}-\vec{\varphi})^\dagger\rangle_{(\vec{\varphi}\vert \vec{d})}\,,
\label{eq:D}
\end{align}
where $\dagger$ denotes adjunction and $\langle \cdot \rangle_{(\vec{\varphi}\vert \vec{d})}$ the expectation value with respect to the posterior probability distribution $P(\vec{\varphi}\vert \vec{d})$. 

In the following sections we will gradually derive the posterior PDF of the physical flux distribution of multiple superimposed photon fluxes given in a data set. 
This will partly follow and build on the existing D$^3$PO algorithm by \cite{Selig:2015rt}.

\subsection{Poissonian likelihood}
\label{subsec:likelihood}
A typical photon count instrument provides us with a data vector $d$  consisting of integer photon counts that are spatially binned into $N_{\text{PIX}}$ pixels. The sky emissivity $\vec{\rho}= \rho(x, E)$, which caused the photon counts, is defined for each continuous sky position $x$ and energy $E$. Since high energy astrophysical spectra cover orders of magnitude in energy, it is convenient to introduce the logarithmic energy coordinate $y=\log(E/E_0)$ with some reference energy $E_0$. The flux is a function of the combined coordinates $z=(x, y)$, such that $\rho(z)= \rho(x,y)$ and lives over the associated domain $\Omega = \mathbb{S} \otimes \mathbb{K}$, i.e. the product of spatial and spectral domains. $\mathbb{S} = \mathcal{S}^2$ if the full sky is treated, $\mathbb{S} = \mathbb{R}^2$ if a patch of the sky is described as in the flat sky approximation, or $\mathbb{S}=\mathbb{R}$ as in the mock examples in this paper, and finally $\mathbb{K} = \mathbb{R}$ as logarithmic energies can be positive and negative. 
The flux $\vec{\rho}$ is a superposition of two morphologically different signal fields, such as a diffuse and point-like flux. Hence
\begin{align}
\vec{\rho}\left(z\right) &= \vec{\rho}\left(z\right)_\text{diffuse} + \vec{\rho}\left(z\right)_\text{point-like}  \notag \\ 
&= \rho_0 \left(e^{s(z)} +e^{u(z)}\right) 
\label{eq:def_rho}
\end{align}
where we introduce the dimensionless fields $s(z)=: \vec{s}$ and $u(z)=: \vec{u}$, to represent the logarithmic diffuse and point-like source fluxes over the signal domain $\Omega$. Further we introduce the convention that a scalar function is applied to a field value by values on its natural domain, $\left( f\left(s\right)\right)\left(z\right)= f\left(s\left(z\right)\right)$ if $z\in \Omega$. The constant $\rho_0$ is absorbing the physical dimensions of the photon flux. \\

The imaging device observing the celestial photon flux $\vec{\rho}$ is expected to measure a number of events $\vec{\lambda}$, informing us about $\vec{s}$ and $\vec{u}$, as well as other event sources like cosmic ray hits or radioactivity, which we will refer to as background. This dependence of $\vec{\lambda}$ on the sky emissivity $\vec{\rho}$ can be modelled via a linear instrument response operator $\vec{R}_0$ acting on $\vec{\rho}$:  
\begin{align}
 \vec{\lambda} &= \vec{R}_0\vec{\rho}  + \vec{R}'_0 \vec{\rho}' \notag \\
 			&= \vec{R} \left(e^{\vec{s}} +e^{\vec{u}}\right) + \vec{R}' e^{\vec{b}} \,,
\label{eq:def_lambda} 
\end{align}
with $\vec{R}= \vec{R}_0\rho_0$ and $\vec{R}'= \vec{R}'_0\rho'_0$. The response operator $\vec{R}_0$ describes all aspects of the measurement processes which relate the sky brightness to the average photon count. To describe the background counts we further introduce the background event emissivity $\vec{\rho}' = \rho'_0 e^{\vec{b}}$, the background event instrument response $\vec{R}'_0$, as well as the abbreviation $\vec{R}' = \vec{R}'_0 \rho'_0$. The background field is a function of the two coordinates exposure time $t$ and the log-energy $y$, such that $ z' = (t, y)$ and its domain $\Omega'= \mathbb{T} \otimes \mathbb{K}$ with $\mathbb{T} \in \mathbb{R}$.  In our applications, we assume that the observing time interval and the sky are identified 
with each other. In real observations, the sky is often scanned multiple times, providing redundancies in the data, which facilitates the separation of sky and background. The background response $\vec{R}'$ describes the instruments sensitivity to background processes, such as cosmic ray events which produce uninteresting counts and need to be filtered out. For each pixel in the detector we get 
\begin{align}
\lambda_i = &\int_{\Omega} dz\, R_i(z) \left( e^{s(z)} + e^{u(z)}\right) + \int_{\Omega'} dz'\, R'_i(z')e^{b(z')}  \,.
\label{eq:pixel_lambda}
\end{align}
The individual events are assumed to follow a Poisson distribution $\mathcal{P}(\vec{d}_i \vert \vec{\lambda}_i)= \vec{\lambda}_i^{\vec{d}_i} e^{-\vec{\lambda}_i}/(\vec{d}_i!)$ in each data bin. Hence the likelihood of the data vector $\vec{d}= (d_1, \dots, d_{N_{\text{PIX}}})$ given an expected number of photons $\vec{\lambda}$ is a product of statistically independent Poisson processes, 
\begin{align}
P(\vec{d}\vert \vec{\lambda}) = \prod_i \mathcal{P}(d_i\vert \lambda_i) = \prod_i \frac{1}{d_i!}\lambda_i^{d_i}e^{-\lambda_i} \,.
\label{eq:Poisson}
\end{align}
In total, the likelihood of photon count data given two superimposed morphologically different photon fluxes and some background fluxes is consequently described by Eq.~(\ref{eq:pixel_lambda}) and Eq.~(\ref{eq:Poisson}). 
To simplify and clarify the notations in the following, we introduce the three component vector $\vec{\varphi}= (\vec{s}^\dagger, \vec{u}^\dagger,\vec{b}^\dagger)^\dagger$ consisting of the diffuse, point-like and the background flux. Further we introduce the combined response $\vec{\mathcal{R}} = (\vec{R}, \vec{R}, \vec{R}')$. 
This allows us to state the information Hamiltonian, defined as the negative logarithm of $P(\vec{d} \vert \vec{\varphi})$,  compactly as
\begin{align}
H(\vec{d}\vert \vec{\varphi}) &= -\log P(\vec{d} \vert \vec{\varphi}) \notag \\
  &= H_0 + \vec{1}^{\dagger} \vec{\mathcal{R}}e^{\vec{\varphi}} - \vec{d}^{\dagger} \log \left(\vec{\mathcal{R}}e^{\vec{\varphi}} \right) \,, 
\label{eq:Hamiltonian_likelihood}
\end{align}
where we absorbed all terms that are constant in $\vec{\varphi}$ into $H_0$ and introduced the scalar product of concatenated vectors
\begin{align}
\vec{\varphi}^\dagger \vec{\varphi'} =& \vec{s}^\dagger \vec{s'} + \vec{u}^\dagger\vec{u'} + \vec{b}^\dagger \vec{b'} \notag\\
						     =& \int_\Omega dz \left(\overline{s\left(z\right)}s'\left(z\right) + \overline{u\left(z\right)}u'\left(z\right)\right) + \int_{\Omega'} dz'\, \overline{b\left(z\right)}b'\left(z\right)\,.
\label{eq:scalar}
\end{align}
$\vec{1}^{\dagger}$ is a constant data vector being one everywhere on the data space. \\

We note that the likelihood, Eq.~(\ref{eq:Poisson}), can be accountable for $n$ signals in data space, with
\begin{align}
\vec{\lambda} = \sum_{\mathrm{i}=1}^\mathrm{n}  \mathcal{R}_\mathrm{i} e^{s_\mathrm{i} \left(z\right)}\,.
\end{align}
This would only extend $\vec{\varphi}$ and $\vec{\mathcal{R}}$ while Eq.~(\ref{eq:Hamiltonian_likelihood}) stays untouched. For this reason, D$^4$PO can reconstruct $n$ such fields, each living over its own space, and all contributing to the expected counts through an individual response function.

\subsection{Prior assumptions}
\label{subsec:prior}
As we are seeking to decompose the photon counts into background flux and two sky flux components, prior PDFs are taken into consideration. Prior knowledge is important to incorporate in the modelling process in order to constrain the degeneracy of the 
inverse problem, since without prior knowledge one can explain the full data set by the diffuse and the point-like signal alone or even purely by a background or any combination thereof. We introduce priors on the fields $\vec{s}, \vec{u},$ and  $\vec{b}$. 
Priors are employed to describe and implicitly define the typical expected morphology of the three different components: diffuse, point-like and background fluxes. 

\subsubsection{The diffuse component}
\label{sec:Prior_diffuse}
The diffuse photon flux $\rho^{(\vec{s})}= \rho_0 e^{\vec{s}}$ is a strictly positive quantity which might vary over several orders of magnitude. Its morphology may be described by cloud-like and smoothly varying patches on the sky. Hence the diffuse flux shows spatial correlations. Furthermore, we concentrate here on radiation processes due to cosmic ray interactions in the interstellar medium like the inverse Compton scattering and $\pi^0$ production and decay. These processes show smooth, often power-law like spectra, with considerable correlation in the log-energy dimension. According to the principle of maximum entropy the log-normal model can be regarded as a minimalistic description of our a priori knowledge on $\rho^{(\vec{s})}$ \citep{2013PhRvE..87c2136O, 2013arXiv1312.6661K}. The log-normal model has shown to be suitable within various observational \citep{Kitaura21112009, Selig:2015ul,2017arXiv170805702P} and theoretical  \citep{Coles01011991, Sheth01121995, 0004-637X-561-1-22, 2001PASP..113.1009V, 2009ApJ...698L..90N,Selig:2015rt, 2016PhRvE..94a2132P,2016A&amp;A...586A..76J,2016arXiv160504317G,2017arXiv171102955K,2017arXiv170101886B} considerations. Hence, we adopt a multivariate Gaussian distribution as a prior for the logarithmic $\vec{s}$:  
\begin{equation}
\mathcal{G} (\vec{s}, \mathcal{S}) = \frac{1}{\sqrt{2 \pi \vert \mathcal{S}\vert}} \exp\left( -\frac{1}{2}\vec{s}^{\dagger} \mathcal{S}^{-1}\vec{s}\right)
\label{eq:Gaussian_s}
\end{equation}
with a covariance $\mathcal{S}= \langle\vec{s}\vec{s}^{\dagger}\rangle_{(\vec{s} \vert \mathcal{S})}$. The covariance $\mathcal{S}$ describes the strength of spatial correlation in the log-energy $y= \log E/E_0$ and space domain $x$ of $\vec{s}$. As these two correlations need to be modelled individually, we chose the following ansatz
\begin{equation}
\mathcal{S}_{zz'} = \mathcal{X}^{(s)} (\vert x-x'\vert) \mathcal{Y}^{(s)}(\vert y- y'\vert )\,, 
\label{eq:ansatz_S}
\end{equation}
being a direct product of the two correlation functions, $\mathcal{X}^{(s)}_{x,x'} = \mathcal{X}^{(s)}\left(\vert x-x'\vert\right)$ and $\mathcal{Y}^{(s)}_{yy'} = \mathcal{Y}^{(s)}(\vert y-y'\vert)$, which only depend on the relative differences in position $x$ and log-energy $y$. This is equivalent to the assumption that $\vec{s}$ is statistically homogeneous and isotropic on the celestial sphere and statistically homogeneous in the log-energy space. Statistical homogeneity in log-energy models the fact that typical high energy astrophysics spectra exhibit similar features on a log-log perspective. Thanks to the assumed statistical homogeneity we can find a diagonal representation of $\mathcal{X}$ and $\mathcal{Y}$ in their harmonic bases, such that 
\begin{align}
\mathcal{X}&= \sum_{l} e^{\tau_\mathcal{X}(l)} \mathbb{P}_l \,, \quad \text{and}\\
\mathcal{Y}&= \sum_{k} e^{\tau_\mathcal{Y}(k)} \mathbb{P}_k  \,.
\label{eq:def_X_Y}
\end{align}
Here $\tau_{\mathcal{X}}(l)$ and $\tau_{\mathcal{Y}}(k)$ are spectral parameters determining the logarithmic power spectra in the spatial and spectral domain, respectively, according to the chosen harmonic basis $(l,k)$ for each corresponding harmonic space. $\mathbb{P}$ is a projection operator onto spectral bands. We assume a similar level of variance for the fields within each of these bands separately. The projection operator is given by: 
\begin{equation}
\mathbb{P}_k \equiv \sum_{k'\in b_k} F_{k'} F_{k'}^{\dagger} \,,
\label{eq:def_projection_operator_k}
\end{equation}
where $F_{k'y}= e^{ik'y}$ is the Fourier basis (or spherical harmonic basis if the subdomain is $\mathcal{S}^2$) and $b_k$ denotes the set of Fourier modes belonging to the corresponding Fourier band $k$. 
The inverses of the two covariances are 
\begin{align}
\mathcal{X}^{-1}&= \sum_{l} e^{-\tau_\mathcal{X}(l)} \mathbb{P}_l \,,\\
\mathcal{Y}^{-1}&= \sum_{k} e^{-\tau_\mathcal{Y}(k)} \mathbb{P}_k  \,.
\label{eq:def_X_Y_inverse}
\end{align}

The spectral parameters $\tau_{\mathcal{X}}$ and $\tau_{\mathcal{Y}}$ are in general unknown a priori. These spectral parameters are reconstructed from the same data as the signal, introducing another prior for their covariances.\\

In the following paragraphs we will introduce two constraints on the spectral parameters $\tau$ . These hyperpriors on the prior itself lead to a hierarchical parameter model. To shorten and clarify notations we will only discuss $\tau_{\mathcal{X}}(l)$ in full detail; the expressions for $\tau_{\mathcal{Y}}(k)$ are analogous. \\

Moreover, $\vec{s}$ may not only depend on location and energy but also on further parameters such as time etc. In this case the covariance $\mathcal{S}$ extends as follows:
\begin{align}
\mathcal{S}_{zz'} &= \prod_\mathrm{i} \mathcal{X}_\mathrm{i} \left(\vert x_\mathrm{i} - x_\mathrm{i}' \vert \right) \quad \mathrm{with} \label{eq:ex_domain}
 \\
z &= \left(x_1, x_2, \dots, x_n\right)^\dagger \notag \\
z'&= \left(x_1', x_2', \dots, x_n'\right)^\dagger \notag\,.
\end{align}
Equation~(\ref{eq:ex_domain}) leads to multiple spectral parameters that need to be inferred from the data if not known a priori. D$^4$PO is also prepared to handle such cases. 

\subsubsection{Unknown magnitude of the power spectrum}
\label{sec:Prior_magnitude}
As the spectral parameters $\tau_\mathcal{X}(l)$ might vary over several orders of magnitude, this demands for a logarithmically uniform prior for each element of the power spectrum and in consequence for a uniform prior $P_{\text{un}}$ for each spectral parameter $\tau_\mathcal{X}(l)$. Following the work of \cite{2011PhRvD..83j5014E, 2010PhRvE..82e1112E} and \cite{Selig:2015rt} we initially assume inverse-Gamma distributions for each individual element,
\begin{equation}
P_{\text{un}}(e^\tau \vert \alpha_l, q_l) = \prod_l \frac{q_l^{\alpha_l-1}}{\Gamma(\alpha_l -1)} e^{-(\alpha_l \tau_l + q_l e^{-\tau_l})}\, 
\label{eq:inverse_gamma}
\end{equation}
and hence, 
\begin{align}
P_{\text{un}}\left( \tau \vert \alpha_l,\,  q_l\right) &= \prod_l \frac{q_l^{\alpha_l-1}}{\Gamma\left(\alpha_l-1\right)}\notag \\
 &\quad \times e^{-\left(\alpha_l \tau_l + q_l e^{-\tau_l}\right)}
  \left \vert \frac{de^{\tau_k}}{d \tau_l}  \right \vert \, ,
 \label{eq:inverse_gamma_tau}
 \end{align}
where $\alpha_l$ and $q_l$ denote shape and scale parameters for the spectral hyperpriors, and $\Gamma$ the Gamma function. 
The form of Eq.~(\ref{eq:inverse_gamma_tau}) shows that for $\alpha_l \rightarrow 1$ and $q_l \rightarrow 0 , \forall l>0,$ the inverse gamma distribution becomes asymptotically flat on a logarithmic scale and therefore for these parameter values do not provide any constraints on the magnitudes of $\tau$. 

\subsubsection{Smoothness of power spectrum}
\label{sec:Prior_smoothness}
Up to now we have only treated each element of the power spectrum separately, permitting the power to change strongly as a function of scale $l$ (and $k$). However, similar spatial (or energetic) scales should exhibit similar amounts of variance for most astrophysical emission processes in the interstellar medium. Thus we assume that the power spectrum is smooth on a logarithmic scale of $l$ and $k$, respectively.  
In accordance with \cite{2011PhRvD..83j5014E} and \cite{2013PhRvE..87c2136O} this can be enforced by introducing a smoothness enforcing prior $P_{\text{sm}}$
\begin{align}
P_{\text{sm}} (\tau \vert \sigma) &\propto \exp \left(-\frac{1}{2 \sigma^{2}} \int d (\log l)\, \left( \frac{\partial^2 \tau_l}{\partial \left( \log l\right)}\right)^{2}\right) \notag \\
&\propto \exp \left(-\frac{1}{2} \tau^{\dagger} \vec{T} \tau\right) \,,
\end{align}
which is based on the second logarithmic derivative of the spectral parameters $\tau$. The parameter $\sigma$ specifies the expected roughness of the spectrum. In the limit $\sigma \rightarrow \infty$ spectral roughness is not suppressed in contrast to $\sigma \rightarrow 0$ which enforces a smooth power-law power spectrum. For a more detailed explanation and demonstration of the influence of $\sigma$ on the reconstruction of $\tau$ we refer to \cite{2017arXiv170805702P}.\\
 
In total the resulting priors for the diffuse flux field are determined by the spectral parameters $\tau_{\mathcal{X^{(\vec{s})}}}$ for spatial correlation and $\tau_{\mathcal{Y^{(\vec{s})}}}$ for correlations in the log-energy domain. These spectral parameters are constrained by  the product of the priors discussed above
\begin{align}
P\left(\tau_i\vert \alpha_i, q_i, \sigma_i\right) =\, & P_{\text{sm}} \left(\tau_i \vert \sigma_i\right) P_\text{un} \left(\tau_i \vert \alpha_i, q_i\right) \,, 
\end{align}
with $i \in \{\mathcal{X}^{(\vec{s})}, \mathcal{Y}^{(\vec{s})}\}$.

\subsubsection{The point-like component}
The photon flux contributions of neighbouring point-like sources in the image space can be assumed to be statistically independent of each other if we ignore knowledge on source clustering. Consequently, statistically independent priors for the photon flux contribution of each point-like source are introduced in the following. 

As the point-like flux $\rho^{(\vec{u})} = \rho_0 e^{\vec{u}}$ is also a strictly positive quantity, we mainly follow the same arguments as for the diffuse flux to derive its prior for correlations within in the log energy domain (Sec.~\ref{sec:Prior_diffuse}). We adopt for the spectral correlation in the energy domain $y$ of $\vec{u}$ a multivariate Gaussian distribution $\mathcal{G}(\vec{u}, \mathcal{Y}^{(u)})$, with $\mathcal{Y}^{(u)}= \langle \vec{u} \vec{u}^\dagger \rangle_{(\vec{u}\vert \vec{d})}$. As we may again assume statistical homogeneity, we can find a diagonal representation of the yet unknown $\mathcal{Y}^{(u)}$, such that
\begin{equation}
\mathcal{Y}^{(u)} = \sum_{k} e^{\tau_{\mathcal{Y}^{(\vec{u})}}(k)} \mathbb{P}_k\,,
\label{eq:def_U}
\end{equation}
with $\mathbb{P}$ being the projection operator according to Eq.~(\ref{eq:def_projection_operator_k}), $\tau_{\mathcal{Y^{(\vec{u})}}}(k)$ being the spectral parameter determining the logarithmic power spectrum of $\mathcal{Y}^{(u)}$. Since $\tau_{\mathcal{Y^{(\vec{u})}}}(k)$ is usually unknown we introduce again a hierarchical parameter model as in Sec.~\ref{sec:Prior_magnitude} and Sec.~\ref{sec:Prior_smoothness}. This gives us 
\begin{align}
P(\tau_{\mathcal{Y}^{(\vec{u})}} \vert \alpha_{\mathcal{Y}^{(\vec{u})}}, q_{\mathcal{Y}^{(\vec{u})}}, \sigma_{\mathcal{Y}^{(\vec{u})}}) =& P_{\text{sm}} (\tau_{\mathcal{Y}^{(\vec{u})}} \vert \sigma_{\mathcal{Y}^{(\vec{u})}}) \notag \\ &\times P_\text{un} (\tau_{\mathcal{Y}^{(\vec{u})}}) \vert \alpha_{\mathcal{Y}^{(\vec{u})}}, q_{\mathcal{Y}^{(\vec{u})}})\,,
\label{eq:prior_point_energy}
\end{align}
with $\alpha_{\mathcal{Y}^{(\vec{u})}}$ and $q_{\mathcal{Y}^{(\vec{u})}}$ being the scale and shape parameters of the inverse gamma distribution and $\sigma_{\mathcal{Y}^{(\vec{u})}}$ the parameter to specify the expected roughness of the spectrum. \\

To derive a suitable prior for the spatial dimension of $\vec{u}$, one may look at the following basic considerations. Let us assume that the universe hosts a homogenous distribution of point sources. Therefore the number of point sources would scale with the observable volume, i.e. with distance cubed. But the apparent brightness of a source is reduced by the spreading of the light rays, i.e. decreases with distances squared. Hence one may expect a power law behaviour between the number of point-like sources and their brightness with a slope of $\beta = 3/2$. As such a power-law is not necessarily normalisable, since it diverges at zero we further impose an exponential cut-off slightly above $0$. This yields an inverse Gamma distribution, which has been shown to be a suitable prior for point-like photon fluxes \citep{2009MNRAS.396..165G, 2009MNRAS.393..681C,2012MNRAS.427.1384C,Selig:2015rt, Selig:2015ul}. The spatial prior for $\vec{u}$ is therefore given by a product of independent inverse-Gamma distributions 
\begin{align}
P\left(\vec{u}_{x}\vert \beta, \eta\right) &= \prod_x \mathcal{I} \left(e^{\vec{u}_x}, \beta, \eta\right)  \left \vert \frac{de^{\vec{u}_x}}{d \vec{u}_x}  \right \vert \\
&\propto \exp \left( -\left( \vec{\beta} -1 \right)^\dagger \vec{u}_x - \vec{\eta}^\dagger e^{\vec{-u}_x}\right)\,,
\end{align} 
where $\vec{\beta} = \beta, \forall x$ and $\vec{\eta} = \eta, \forall x$ are the shape and scale parameters of the inverse gamma distribution $\mathcal{I}$. \\
In total the prior for the point-like flux becomes 
\begin{align}
P(\vec{u}) =&\, P\left(\vec{u} _y\vert \tau_{\mathcal{Y}^{(\vec{u})}}\right) P\left(\tau_{\mathcal{Y}^{(\vec{u})}} \vert \alpha_{\mathcal{Y}^{(\vec{u})}}, q_{\mathcal{Y}^{(\vec{u})}}, \sigma_{\mathcal{Y}^{(\vec{u})}}\right) P\left(\vec{u}_{x}\vert \beta, \eta\right)\,.
\label{eq:Prior_u}
\end{align}

\subsubsection{Modelling the background}

Since the background $\vec{b}$ is also a strictly positive but unknown field, we assume the prior $P(\vec{b})$ to have the same structure as the one for $\vec{s}$, except for the fact that it might be defined over different spaces. Hence we may again follow Sec.~\ref{sec:Prior_diffuse} to build a hierarchical prior model for $P(\vec{b})$. 

\subsubsection{Parameter model}
    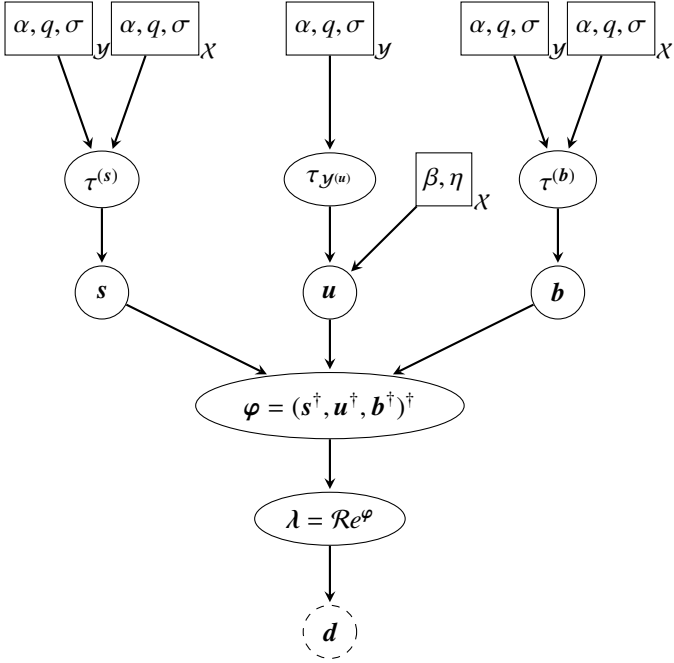
\begin{figure}
    \centering
        \begin{tikzpicture}
            [c/.style={circle,minimum size=2em,text centered,thin},
             r/.style={rectangle,minimum size=2em,text centered,thin},
             rd/.style={ellipse,minimum size=2em,text centered,thin, dashed},
             e/.style={ellipse,minimum size=2em,text centered,thin},
             v/.style={->,shorten >=1pt,>=stealth,thick}, 
             arrow/.style={-latex, shorten >=1ex, shorten <=1ex, bend angle=45}]
            \node(a)at(-3.7,2)[r, draw, fill=white]{$\alpha, q, \sigma$};
            \node(ay)at(-3, 1.7){\tiny{$\mathcal{Y}$}};
            \begin{scope}[on background layer]
            \node(a1)at(-2.3, 2)[r, draw]{$\alpha, q, \sigma$};
 	    \node(ax)at(-1.6, 1.7) {\tiny{$\mathcal{X}$}};      	
             \end{scope}
	    \node(c)at(0,2)[r, draw]{$\alpha, q, \sigma$};
	    \node(cy)at(.7, 1.7){\tiny{$\mathcal{Y}$}};
	    \node(d)at(2.3,2)[r, draw, fill= white]{$\alpha, q, \sigma$};
	    \node(cy)at(3., 1.7){\tiny{$\mathcal{Y}$}};
	    \begin{scope}[on background layer]
            \node(c1)at(3.7, 2)[r, draw]{$\alpha, q, \sigma$};
 	    \node(cx)at(4.4, 1.7) {\tiny{$\mathcal{X}$}};      	
             \end{scope}
             \node(sx)at(1.5, 0)[r, draw]{$\beta, \eta$};
             \node(cx)at(2., -.3) {\tiny{$\mathcal{X}$}};  
	    \node(e)at(-3, 0)[e, draw]{$\tau^{(\vec{s})}$};
	    \node(e1)at(-3,-1.5)[e, draw]{$ \vec{s}$};
	    \node(f)at(0,0)[e, draw]{$\tau_{\mathcal{Y}^{(\vec{u})}}$};
	    \node(f1)at(0, -1.5)[e, draw]{$\vec{u}$};
	    \node(g)at (3, 0)[e, draw]{$\tau^{(\vec{b})}$};
	    \node(g1)at(3, -1.5)[e, draw]{$\vec{b}$}; 
	    \node(h)at (0, -3)[e, draw]{$\vec{\varphi}= (\vec{s}^\dagger, \vec{u}^\dagger, \vec{b}^\dagger)^\dagger$}; 
	    \node(i)at(0, -4.5)[e, draw]{$\vec{\lambda}= \mathcal{R}e^{\vec{\varphi}}$};
	    \node(j)at(0, -6)[rd, draw]{$\vec{d}$};
	    \draw[v](a1)--(e);
	    \draw[v](c1)--(g);
	    \draw[v](sx)--(f1);
	    \draw[v](a)--(e);
	    \draw[v](d)--(g);
	    \draw[v](c)--(f);
	    \draw[v](e)--(e1);
	    \draw[v](g)--(g1);
	    \draw[v](f)--(f1);
	    \draw[v](h)--(i);
	    \draw[v](i)--(j);
	    \draw[v](e1)--(h);
	    \draw[v](f1)--(h);
	    \draw[v](g1)--(h);
        \end{tikzpicture}
        \flushleft
        \caption{Graphical model of the hierarchical Bayesian network introduced in Section~\ref{sec:inference}. Shown are the model parameters $\alpha, q, \sigma, \eta$ and $\beta$, in rectangular boxes, as they have to be specified by the user. The logarithmic spectral parameters $\tau^{(\vec{s})}= (\tau_{\mathcal{X}^{(\vec{s})}}^\dagger, \tau_{\mathcal{Y}^{(\vec{s})}}^\dagger)^\dagger, \tau_{\mathcal{Y}^{(\vec{u})}}^\dagger$ and $\tau^{(\vec{b})}= (\tau^\dagger_{\mathcal{X}^{(\vec{b})}}, \tau^\dagger_{\mathcal{Y}^{(\vec{b})}})^\dagger,$ the diffuse signal field $\vec{\varphi}$, and the expected number of photons $\vec{\lambda}$, are inferred by the algorithm and shown in black solid circles. The observed photon count data, $\vec{d}$, is marked by a dashed circle at the bottom.}
        \label{fig:parameter_model}
\end{figure} 

Figure~\ref{fig:parameter_model} shows the complete hierarchical Bayesian network introduced in Section~\ref{sec:inference}. The goal is to model the data $\vec{d}$  by our quantities $\vec{s}$ and $\vec{u}$ and the background described by $\vec{b}$. The three logarithmic power spectra $\tau^{(\vec{s})}= (\tau_{\mathcal{X}^{(\vec{s})}}^\dagger, \tau_{\mathcal{Y}^{(\vec{s})}}^\dagger)^\dagger$, $\tau_{\mathcal{Y}^{(\vec{u})}}^\dagger$, and $\tau^{(\vec{b})}= (\tau^\dagger_{\mathcal{X}^{(\vec{b})}}, \tau^\dagger_{\mathcal{Y}^{(\vec{b})}})^\dagger,$ can be reconstructed from the data. In case additional information sources are available to further constraint these parameters one may certainly use those and adjust the known logarithmic power spectra accordingly. However, in total the suggested algorithm is steered by the parameters $\alpha, q,$ and $\sigma$, for which we can partly provide canonical values:  $\alpha_{k,i}=\sigma_{k,i}= 1$ $\forall \, k = 0$, $\beta=3/2$,  as well as $\eta, q_{i} \gtrsim 0$ for $i \in \{\mathcal{X}^{(\vec{s})}, \mathcal{Y}^{(\vec{s})}, \mathcal{Y}^{(\vec{u})}, \mathcal{X}^{(\vec{b})}, \mathcal{Y}^{(\vec{b})}\}$. 

\section{The inference}
\label{sec:Inference}
The likelihood constructed in Sec.~\ref{sec:inference} and the prior assumptions for the diffuse, point-like, and background signal field contain together all information available to tackle this inference problem. The resulting posterior PDF is given by
\begin{align}
P(\vec{\varphi}, \vec{\tau} \vert \vec{d}) =&\, \frac{P(\vec{d}\vert \vec{\varphi}, \vec{\tau})}{P(\vec{d})} \notag \\
& \times \prod_{i \in \{\vec{s}, \vec{u}, \vec{b}\}} P\left(\tau_{\mathcal{Y}^{(i)}} \vert \alpha_{\mathcal{Y}^{(i)}}, q_{\mathcal{Y}^{(i)}}, \sigma_{\mathcal{Y}^{(i)}}\right) \notag \\
& \times \prod_{i \in \{\vec{s}, \vec{b}\}} P\left(\tau_{\mathcal{X}^{(i)}} \vert \alpha_{\mathcal{X}^{(i)}}, q_{\mathcal{X}^{(i)}}, \sigma_{\mathcal{X}^{(i)}}\right) \notag \\
& \times P\left(\vec{\varphi}_{\vec{u}_{x}}\vert \beta, \eta\right)
\label{eq:posterior_complete}
\end{align}
where we have introduced $\vec{\tau} = (\tau_{\mathcal{X}^{(\vec{s})}}, \tau_{\mathcal{Y}^{(\vec{s})}}, \tau_{\mathcal{Y}^{(\vec{u})}}, \tau_{\mathcal{X}^{(\vec{b})}}, \tau_{\mathcal{Y}^{(\vec{b})}})$. 

In an ideal case with unlimited computational power, we would now calculate the mean and its variances according to Eq.~(\ref{eq:m}) and (\ref{eq:D}) for $\vec{\varphi}$, by integrating over all possible combinations of $\vec{\varphi}$ and $\vec{\tau}$. This would also provide us with all spectral parameters $\vec{\tau}$. But due to the complexity of the posterior probability distribution this is not worth pursuing.  

We are relying on numerical approaches. Phase space sampling techniques like Markov chain Monte Carlo methods \citep{10.2307/2280232, 2004astro.ph..1623W, 2010MNRAS.407...29J, 2010MNRAS.406...60J, 2013ApJ...779...15J} are hardly applicable to our inference problem due to the extremely large phase space to be sampled over. Consequently we have to find suitable approximations to tackle the problem.\\
In Sec.~\ref{subsec:map} we introduce a Maximum a posteriori approach to solve the inference problem and in Sec.~\ref{subsec:physical} we show how to obtain physical solutions from this approximation.

\subsection{Maximum a posteriori}
\label{subsec:map}
In case the posterior distribution is single peaked and symmetric, its maximum and mean coincide. At least in first order approximation this holds for Eq.~(\ref{eq:posterior_complete}), which allows us to use the so called maximum a posteriori (MAP) approach. This method can be enforced by introducing either a $\delta-$function at the posterior's mode, 
\begin{equation}
\langle \vec{\varphi}\rangle_{(\vec{\varphi}\vert \vec{d})} \overset{\text{MAP-}\delta}{\approx} \int \mathcal{D}\vec{\varphi} \, \vec{\varphi}\,\delta(\vec{\varphi}-\vec{\varphi}_{\text{mode}})
\label{eq:map_d}
\end{equation}
or by using a Laplace approximation, with its uncertainty covariance $D$, estimated from the curvature around the maximum. A field expectation value is given by
\begin{equation}
\langle f\left(\vec{\varphi}\right)\rangle_{(\vec{\varphi}\vert \vec{d})} \overset{\text{MAP-}\mathcal{G}}{\approx} \int \mathcal{D}\vec{\varphi} \, f\left(\vec{\varphi}\right)\,\mathcal{G} \left(\vec{\varphi}-\vec{\varphi}_{\text{mode}}, D\right)\,.
\label{eq:map_G}
\end{equation} \\

In either case we are required to find the mode which is the maximum of the posterior distribution \eqref{eq:posterior_complete}.
Rather than maximizing the full posterior, it is convenient to minimize the information Hamiltonian, defined by its negative logarithm
\begin{align}
H(\vec{\varphi}, \vec{\tau} \vert \vec{d}) =& - \log P(\vec{\varphi}, \vec{\tau} \vert \vec{d}) \notag \\
  =& \, H_0 + \vec{1}^{\dagger} \vec{\mathcal{R}}e^{\vec{\varphi}} - \vec{d}^{\dagger} \log \left[\vec{\mathcal{R}}e^{\vec{\varphi}} \right] \notag \\
&+ \frac{1}{2}  \left[ \log (\det \Phi) + \vec{\varphi}^\dagger \Phi^{-1} \vec{\varphi} \right] \notag \\ 
& + \sum_{i \in \mathcal{I}} (\vec{\alpha}_i -1)^\dagger \tau_i + \vec{q}_i^\dagger e^{-\tau_i} + \frac{1}{2} \tau_i^\dagger \vec{T}_{i} \tau_i   \notag \\
& + \left(\vec{\beta} - 1\right)^\dagger\vec{\varphi}_{\vec{u}_x} + \vec{\eta}^\dagger e^{-\vec{\varphi}_{\vec{u}_x}}
\label{eq:Hamiltonian}
\end{align}
with  $\Phi = \diag (\mathcal{S}, \mathcal{U}, \mathcal{B})^\mathrm{T}$ and $\mathcal{I}  \in \{ \mathcal{X}^{(\vec{s})}, \mathcal{Y}^{(\vec{s})}, \mathcal{Y}^{\vec{(u)}}, \mathcal{X}^{(\vec{b})}, \mathcal{Y}^{(\vec{b})}\}$. We have absorbed all terms that are constant in $\vec{\varphi}$ and $\vec{\tau}$ into $H_0$. 
The MAP-ansatz seeks for the minimum of  Eq.~(\ref{eq:Hamiltonian}), which is equivalent to maximizing the posterior Eq.~(\ref{eq:posterior_complete}). This minimum can be found by taking the first partial derivatives of Eq.~(\ref{eq:Hamiltonian}) with respect to all components of $\vec{\varphi}$ and $\vec{\tau}$, respectively and equalling them to zero. The resulting filtering formulas for the diffuse and point-like flux read as
\begin{align}
\left .\frac{\partial H}{\partial \vec{\varphi}} \right\vert_{\text{min}} =& \left( 1- \frac{\vec{d}}{\vec{l}}\right)^\dagger \vec{R}* e^{\vec{\varphi}} + \Phi^{(*)-1}\vec{\varphi} \overset{!}{=} 0\notag \\
\left .\frac{\partial H}{\partial \vec{\varphi}_{\vec{u}_x}} \right\vert_{\text{min}} = & \left .\frac{\partial H}{\partial \vec{\varphi}} \right\vert_{\text{min}} + \left(\vec{\beta} -1\right) - \vec{\eta} * e^{-\vec{\varphi}_{\vec{u}_x}}\overset{!}{=} 0
 \label{eq:map_varphi_grad}
\end{align}
with
\begin{align}
\vec{l} &= \vec{\mathcal{R}}e^{\vec{\varphi}} \,, \label{eq:Hamiltonian_l}\\
\Phi^{(*)} &= \diag \left(\mathcal{S}^{(*)}, \mathcal{U}^{(*)}, \mathcal{B}^{(*)}\right)^\mathrm{T}  \\
\mathcal{S}^{(*)} &= \sum_k e^{\tau^{(*)}_{\mathcal{Y}^{(\vec{s})}}(k)} \mathbb{P}_k \sum_l e^{\tau^{(*)}_{\mathcal{X}^{(\vec{s})}}(l)} \mathbb{P}_l\,,\\
\mathcal{U}^{(*)} &= \sum_k e^{\tau^{(*)}_{\mathcal{Y}^{(\vec{u})}}(k)} \mathbb{P}_k\,, \\
\mathcal{B}^{(*)} &= \sum_k e^{\tau^{(*)}_{\mathcal{Y}^{(\vec{b})}}(k)} \mathbb{P}_k \sum_k e^{\tau^{(*)}_{\mathcal{X}^{(\vec{b})}}(k)} \mathbb{P}_k\,.
\end{align}
By $*$ and $\frac{\vec{d}}{\vec{l}}$ we refer to component wise multiplication and division, respectively. 
The filtering formulas for the power spectra, which are also derived by taking the first partial derivatives with respect to the components of $\vec{\tau}$ read as
\begin{align}
e^{\tau_{\mathcal{Y}^{(\vec{i})}}}  &= \frac{q_{\mathcal{Y}^{(\vec{i})}} + \frac{1}{2} \left( \Tr\left [\vec{i}\vec{i}^\dagger \mathbb{P}^\dagger\mathcal{X}^{-1}\right]\right)}{\alpha_{\mathcal{Y}^{(\vec{i})}} -1 + \frac{1}{2} \left( \Tr \left [ \mathbb{P} \mathbb{P}^\dagger \mathcal{X}\right]\right)_k  + \vec{T}_{\mathcal{Y}^{(i)}}\tau_{\mathcal{Y}^{(\vec{i})}}}\,, \label{eq:map_i}\\ 
\text{and} \notag \\
e^{\tau_{\mathcal{X}^{(\vec{j})}}} &= \frac{q_{\mathcal{X}^{(\vec{j})}} + \frac{1}{2} \left( \Tr\left [\vec{j}\vec{j}^\dagger \mathbb{P}^\dagger\mathcal{Y}^{-1}\right]\right)}{\alpha_{\mathcal{X}^{(\vec{j})}} -1 + \frac{1}{2} \left( \Tr \left [ \mathbb{P}\mathbb{P}^\dagger \mathcal{Y}\right]\right)_l  + \vec{T}_{\mathcal{X}^{(j)}}\tau_{\mathcal{X}^{(\vec{j})}}}\,,\label{eq:map_j}
\end{align}
with $ i \in \{\vec{s}, \vec{u}, \vec{b} \}$ and $j \in \{\vec{s}, \vec{b}\}$. 
These filtering formulas for the spectral parameters $\vec{\tau}$ are in accordance with \cite{2011PhRvD..83j5014E, 2013PhRvE..87c2136O}. Unfortunately the Eq.~(\ref{eq:map_varphi_grad}) to (\ref{eq:map_j}) lead to eight implicit equations rather than one explicit.  Hence these equations need to be solved by an iterative minimization of Eq.~(\ref{eq:Hamiltonian}) using a minimisation algorithm such as steepest descent. 
The second derivative of the Hamiltonian, i.e. the Hessian around the minimum, may serve as a first order approximation of the uncertainty covariance,
\begin{align}
\left . \frac{\partial^2 H}{\partial \vec{\varphi}\partial\vec{\varphi}^\dagger} \right\vert_{\text{min}} \approx D^{\left(\vec{\varphi}\right)-1}\,.
\end{align}
A detailed derivation and closed form of $D^{\left(\vec{\varphi}\right)-1}$ can be found in \Cref{app:curvature}. \\

It has been shown \citep{2011PhRvD..83j5014E} that MAP-estimating of a field and its power spectrum simultaneously is suboptimal. The reason is that the joined posterior is far from being symmetric and exhibits long tails in directions correlated in the field and in its spectrum. The resulting scheme can strongly underestimate the field variance in low signal-to-noise situations. To overcome this difficulty, a number of approaches have been pursued like renormalization techniques. They lead to improved schemes which are closely related to each other. The most transparent of these approaches is the mean field approximation that involves the construction and minimization of an action, the Gibbs free energy \citep{2010PhRvE..82e1112E}. A detailed derivation of this approach can be found in \Cref{app:gibbs}. \\

The filtering formulas in a Gibbs free energy approach for $\vec{\varphi}$ only change within the exponent, while the one for $\vec{\tau}$ yields
\begin{align}
e^{\tau_{\mathcal{Y}^{(\vec{i})}}}  &= \frac{q_{\mathcal{Y}^{(\vec{i})}} + \frac{1}{2} \left( \Tr\left [\left(\vec{i}\vec{i}^\dagger + D^{\left(j\right)}\right) \mathbb{P}^\dagger\mathcal{X}^{-1}\right]\right)}{\alpha_{\mathcal{Y}^{(\vec{i})}} -1 + \frac{1}{2} \left( \Tr \left [ \mathbb{P} \mathbb{P}^\dagger \mathcal{X}\right]\right)_k  + \vec{T}_{\mathcal{Y}^{(i)}}\vec{t}_{\mathcal{Y}^{(\vec{i})}}}\,, \label{eq:Gibbs_tau_i}\\ 
\text{and} \notag \\
e^{\tau_{\mathcal{X}^{(\vec{j})}}} &= \frac{q_{\mathcal{X}^{(\vec{j})}} + \frac{1}{2} \left( \Tr\left [\left(\vec{j}\vec{j}^\dagger + D^{\left(j\right)}\right) \mathbb{P}^\dagger\mathcal{Y}^{-1}\right]\right)}{\alpha_{\mathcal{X}^{(\vec{j})}} -1 + \frac{1}{2} \left( \Tr \left [ \mathbb{P}\mathbb{P}^\dagger \mathcal{Y}\right]\right)_l  + \vec{T}_{\mathcal{X}^{(j)}}\vec{t}_{\mathcal{X}^{(\vec{j})}}}\,,\label{eq:Gibbs_tau_j}
\end{align}
with $ i \in \{\vec{s}, \vec{u}, \vec{b} \}$ and $j \in \{\vec{s}, \vec{b}\}$. 
These formulas are in close accordance with the critical filtering technique \citep{2013PhRvE..87c2136O}. Equations~(\ref{eq:Gibbs_tau_i}) and (\ref{eq:Gibbs_tau_j}) give a reconstruction of the spectral parameters of a field which does not necessarily need to be statistically isotropic and homogeneous over its combined domains. The appearing correction term $D$ in the trace term of the numerator of Eq.~(\ref{eq:Gibbs_tau_j}) and (\ref{eq:Gibbs_tau_i}) compared to the MAP-solutions Eq.~(\ref{eq:map_i}) and (\ref{eq:map_j}) is positive definite. Hence it introduces a positive contribution to the logarithmic power spectrum and therefore lowers a potential perception threshold \citep{2011PhRvD..83j5014E}. 

\subsection{The physical solution}
\label{subsec:physical}
All previously described methods only recover logarithmic fluxes, but the actual quantities of interest are the physical fluxes $\vec{\rho}$. These physical fluxes can be calculated employing the following approximation: 
\begin{align}
\left \langle \vec{\rho}\right\rangle  \overset{\text{MAP}-\delta}{\approx}& \left \langle \vec{\rho}\right\rangle_\delta = \rho_0 e^{m_{\text{mode}}} \label{eq:map}\\
 \overset{\text{MAP-G}}{\approx}& \left \langle \vec{\rho}\right\rangle_\mathcal{G} = \rho_0 e^{m_{\text{mode}}+ \frac{1}{2}D_{\text{mode}}}\,.
 \label{eq:physical_sol}
\end{align}
The uncertainty of the reconstructed fields may be approximated by
\begin{align}
\sigma_{\mathcal{G}}^2 = \left\langle\vec{\rho}^2\right\rangle_{\mathcal{G}}-\Big\langle\vec{\rho}\Big\rangle^2_{\mathcal{G}} \overset{\text{MAP-G}}{\approx}& \Big\langle\vec{\rho}\Big\rangle_{\mathcal{G}}^2 \left( e^D -1\right) \,,
\label{eq:physical_uncert}
\end{align}
with its square root being the relative uncertainty. \\
The full Gibbs approach described in Sec.~\ref{subsec:Gibbs} would require to know $D^{\left(\varphi\right)}$ at all times during the minimisation of Eq.~(\ref{eq:full_Gibbs}), and is numerically not feasible. We only consider the MAP-$\mathcal{G}$ approach here to infer the signal fields. D$^{3}$PO has shown that such an approach does produce accurate signal field reconstructions. For the reconstruction of the spectral parameters $\vec{\tau}$, we use the full Gibbs approach as given by Eq.~(\ref{eq:Gibbs_tau_i}) and (\ref{eq:Gibbs_tau_j}), as these take higher order correction terms into account. \\ 
It must be noted that the mode approximation only holds for strictly convex problems and can perform poorly if this property does not hold. The precise form of the posterior is neither analytically nor numerically fully accessible due to the potentially extremely large phase space of the degrees of freedom. However in first order approximation, the posterior Hamiltonian may be assumed to be convex close to its minimum.  

\section{The inference algorithm}
\label{sec:Algorithm}
To denoise, deconvolve, and decompose photon observations, while simultaneously learning the statistical properties of the fields, i.e. their power spectra, is a highly relevant but non-trivial task. \\
The derived information Hamiltonian Eq.~(\ref{eq:Hamiltonian}) and Gibbs free energy Eq.~(\ref{eq:full_Gibbs}) are scalar quantities defined over a potentially huge phase space of $\vec{\varphi}$ and $\vec{\tau}$. Even within an ideal measurement scenario the inference has to estimate three numbers plus the spectral parameters for each location and energy of the field from just one data value. Hence the inference can become highly degenerate if the data or priors do not sufficiently constrain the reconstruction. Such a scenario would likely lead to multiple local minima in a non-convex manifold of the landscape of the information Hamiltonian and Gibbs free energy, respectively. In total the complexity of the inference has its main roots at the non-linear coupling between the individual fields and spectral parameters to be inferred. \\
After numerous numerical tests we can propose an iterative optimisation scheme, which divides the global minimisation into multiple, more easily solvable subsets. By now, the following guide has given the best results:
\begin{enumerate}[noitemsep,nolistsep]
\item Initialise the algorithm with naive starting values, i.e. $\vec{\varphi}=0$ and $e^{\vec{\tau}_k} \propto k^{-4}$. If more profound knowledge is at hand you may certainly use this to construct a suitable initial field prior in order to speed up the inference and to overcome the outlined issues about the non-convexity of the minimisation. 
\item Optimise the diffuse signal field, by minimising the information Hamiltonian Eq.~(\ref{eq:Hamiltonian}) or the Gibbs free energy Eq.~(\ref{eq:full_Gibbs}), respectively, by a method of your choice. As the gradients are always analytically accessible, we recommend using methods which make use of this, such as steepest descent. 
\item Optimise the point-like and background signal field, accordingly to the diffuse signal field in the step before. 
\item Update both spectral parameters of the diffuse flux field by again minimising Eq.~(\ref{eq:Hamiltonian}) and Eq.~(\ref{eq:full_Gibbs}), respectively, with respect to $\tau_{\mathcal{X}}^{\left(s\right)}$ and $\tau_{\mathcal{Y}}^{\left(s\right)}$. This optimisation step may be executed via a quasi Newton method. 
\item Optimise the diffuse flux field, analogous to step two. 
\item Optimise the spectral parameters and maps of the point-like and background signal field according to step four and five, analog to the diffuse signal field. 
\item Iterate between the steps four to six, until you have reached a global optimum. This may take a couple cycles. In order to get rough estimates of the signal in early cycles it is not necessary to let each minimisation run until it has reached its global desired convergence criteria. These may be retightened gradually as one gets closer to the global minimum. 
\item At the reached minimum, calculate $D^{\left(\varphi\right)}$ in order to obtain the physical fluxes Eq.~(\ref{eq:physical_sol}).
\end{enumerate}
It must be noted that the outlined iterative minimisation scheme has proven in multiple numerical tests to lead to the global minimum. However due to the extremely large phase space it is almost impossible to judge whether the algorithm has truly converged to the global optimum in case one works with real astrophysical data sets. Nevertheless this should not matter too much, in case the local minimum and the global optimum are close to each other and therefore do not differ substantially. 

\section{An inference example}
\label{sec:mock_example}
In order to demonstrate the performance of the inference algorithm we apply the D$^4$P0 algorithm to a realistic but simulated astrophysical data set. 
In this mock example the algorithm is required to reconstruct the diffuse, the point-like, and the background fluxes. Additionally we request to infer all statistical properties of the diffuse flux, i.e. $\tau_{\mathcal{X}}^{\left(s\right)}$ and $\tau_{\mathcal{Y}}^{\left(s\right)}$, and $\tau_{\mathcal{Y}}^{\left(u\right)}$ of the point flux. The statistical properties of the background radiation are assumed to be known, otherwise the inference problem would be completely degenerate as no prior information would separate the background from the diffuse flux. 
The mock data set originates from a hypothetical observation with a field of view of $350 \times 350$ pixels and a resolution of $0.005 \times 0.01$ [a.u.]. 
All signal fields were drawn from Gaussian random fields with different correlation structures. The functional form of all correlation structures is 
\begin{align}
e^{\vec{\tau}\left(k\right)} =& \frac{\theta^2 \kappa }{\left(1+ \left(\frac{2 \pi k \kappa }{4}\right)^2\right)^2}\, ,
\label{eq:correlation}
\end{align}
but the correlation length $\kappa$ and the variance $\theta$ differ for each field and their sub-domains. The chosen parameters are given by  \Cref{table:parameters_corr}.
\begin{table}
\caption{Parameters to define correlation structure of $\vec{s}$, $\vec{u}$, and $\vec{b}$}              
\label{table:parameters_corr}      
\centering                                      
\begin{tabular}{c | c c c c c }          
& $e^{\tau_\mathcal{X}^{\left(\vec{s}\right)}}$ &$e^{\tau_\mathcal{Y}^{\left(\vec{s}\right)}}$ & $e^{\tau_\mathcal{Y}^{\left(\vec{u}\right)}}$& $e^{\tau_\mathcal{X}^{\left(\vec{b}\right)}}$& $e^{\tau_\mathcal{Y}^{\left(\vec{b}\right)}}$ \\   
\hline\hline
$\theta$ & $4.5$ & $0.3$ & $0.1$ & $0.75$ & $2.0$ \\ 
$\kappa$ & $2.5$ & $0.2$ & $2.5$ & $0.005$ & $0.014$
\end{tabular}
\end{table}

\begin{figure*}[]
\centering
\includegraphics[width=.9625\textwidth]{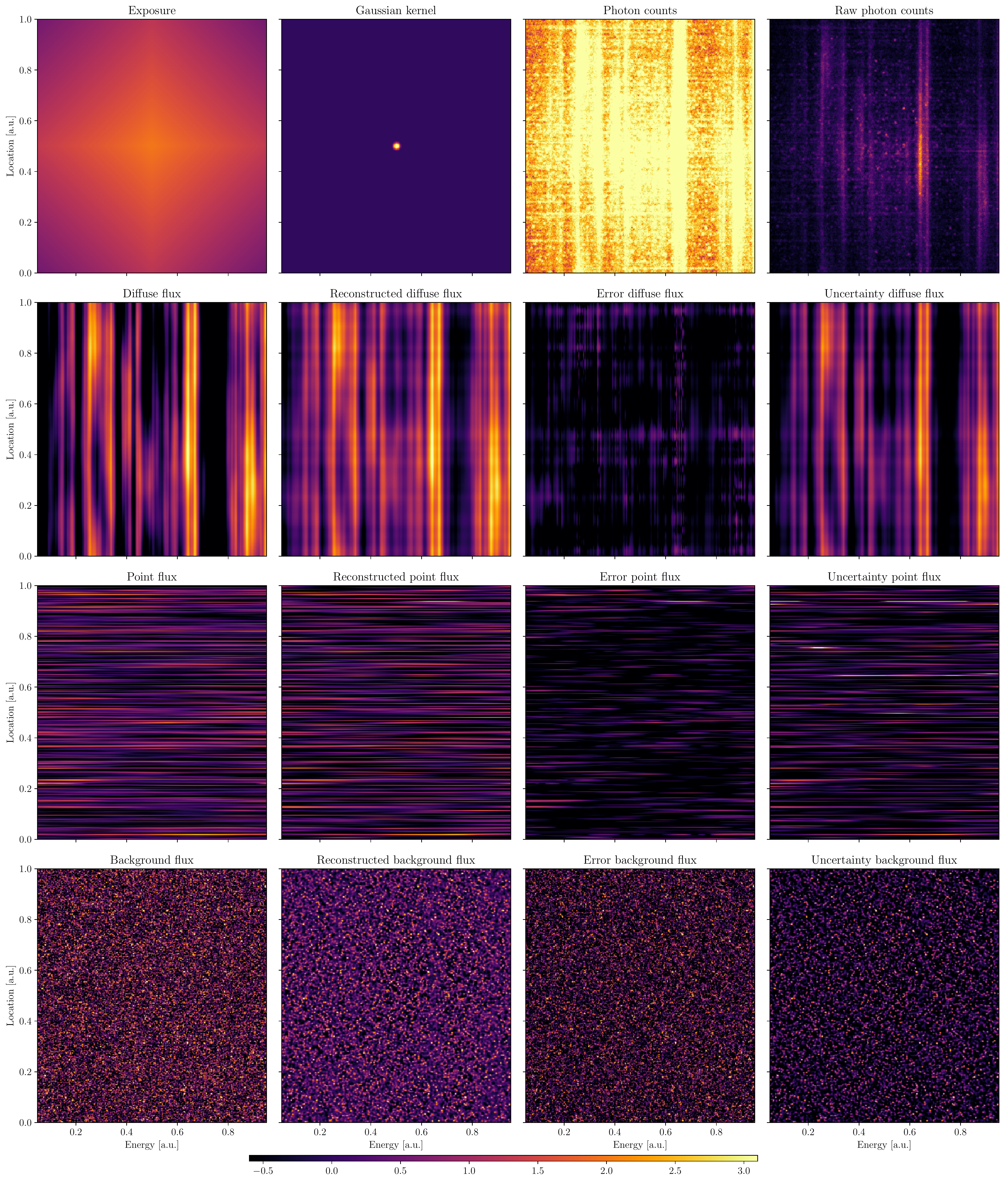}
\caption{Demonstration of the capabilities of D$^4$PO based on a simulated but realistic data set gathered from a potential astrophysical high energy telescope. The spatial dimension is orientated vertically and the spectral/energy dimension horizontally. Point-like sources appear therefore as the horizontal lines. All fields are living over a regular grid of $350 \times 350$ pixels. The top panel shows the assumed instrument's exposure map, its Gaussian convolution kernel and the obtained data set, once on logarithmic scale and once the raw photon counts. For each photon count we obtain spectral and spatial information. The panels below display the diffuse, point-like, and background flux. For all signal fields we show the ground truth on the left hand side ($\vec{s}_{\text{min}=-3.80}^{\text{max}=3.58}, \vec{u}_{\text{min}=-7.80}^{\text{max}=2.73}, \vec{b}_{\text{min}=-5.22}^{\text{max}=4.91}$), followed by its reconstruction ($\vec{s_\text{rec}}_{\text{min}=-.93}^{\text{max}=3.61}, \vec{u_\text{rec}}_{\text{min}=-2.21}^{\text{max}=3.67}, \vec{b_\text{rec}}_{\text{min}=-1.24}^{\text{max}=4.875}$), the error ($\vec{s_\text{error}}_{\text{min}=-1.06}^{\text{max}=1.46}, \vec{u_\text{error}}_{\text{min}=-1.36}^{\text{max}=3.64}, \vec{b_\text{error}}_{\text{min}=-1.37}^{\text{max}=3.70}$) and the logarithmic uncertainty of the reconstruction ($\vec{s_\text{unc}}_{\text{min}=-1.43}^{\text{max}=3.11}, \vec{u_\text{unc}}_{\text{min}=-2.72}^{\text{max}=4.93}, \vec{b_\text{unc}}_{\text{min}=-1.74}^{\text{max}=4.38}$). For illustration purposes all fluxes are on logarithmic scale and clipped between $-0.6$ and $3.1$, except the `Raw photon counts' ($\vec{d}_{\text{min}=0}^{\text{max}=175}$) which are shown on their native scale.}
\label{fig:sub_maps}
\end{figure*}

The assumed instrument's response incorporates a convolution with a Gaussian-like PSF with a FWHM of two times the pixelation size in each direction and an inhomogeneous exposure. The logarithmic exposure, the logarithmic PSF, the logarithmic photon counts, as well as the raw photon counts are shown in the top panel of Fig.~\ref{fig:sub_maps}. The obtained data set provides spatial and spectral information about each observed photon. \\

Further rows of Fig.~\ref{fig:sub_maps} show all signal fields in terms of logarithmic fluxes, i.e. $\vec{s}, \vec{u}$, and $\vec{b}$, all depending on location on the vertical axis and energy on the horizontal axis. 
For each field we show the ground truth, i.e. the drawn Gaussian random field, its reconstruction, the error (or residual) between reconstruction and truth flux, i.e. $\left\vert \vec{\rho}^{\mathrm{rec}}- \vec{\rho}^{\mathrm{truth}}\right \vert$, as well as the uncertainty $\sigma_{\mathcal{G}}$ of the reconstruction provided by D$^4$PO, according to Eq.~(\ref{eq:physical_uncert}).  For the reconstruction we used the following parameter setup, $\alpha_\mathrm{i} = 1, q_\mathrm{i}=10^{-12}, \sigma_\mathrm{i}=1, \,  \mathrm{i} 
 \in \{ \mathcal{X}^{(\vec{s})}, \mathcal{Y}^{(\vec{s})}, \mathcal{Y}^{\vec{(u)}}\},\, \beta=\frac{3}{2}$, and  $\eta=10^{-4}$ in a MAP-$\mathcal{G}$ approach as this has proven to give the best results within a reasonable amount of computing time \citep{Selig:2015rt}. \\
 Looking more closely at the diffuse flux field, the original and its reconstruction are in good agreement. The strongest deviation may be found in regions with low amplitudes, which is not surprising as we are using an exponential ansatz to enforce positivity for all our fields. Hence small errors in $\vec{s} \rightarrow \left(1 \pm \epsilon\right)\vec{s}$ factorise in the physical photon flux field, $\vec{\rho}^{\left(\vec{s}\right)} \rightarrow e^{\vec{s}} e^{\pm\epsilon}$, that scales exponentially with the amplitude of the diffuse flux field. Further, in almost all regions the absolute error shows that the reconstruction is in very good agreement with the original one. Only in areas with a relatively weak point flux and a rather strong diffuse flux the decomposition seems to run into a fundamental problem, as the priors and the likelihood can no longer break the degeneracies between the different sources. \\
\begin{figure*}[]
\centering
	\begin{subfigure}{6cm}
	\centering 
		\includegraphics[width=\textwidth]{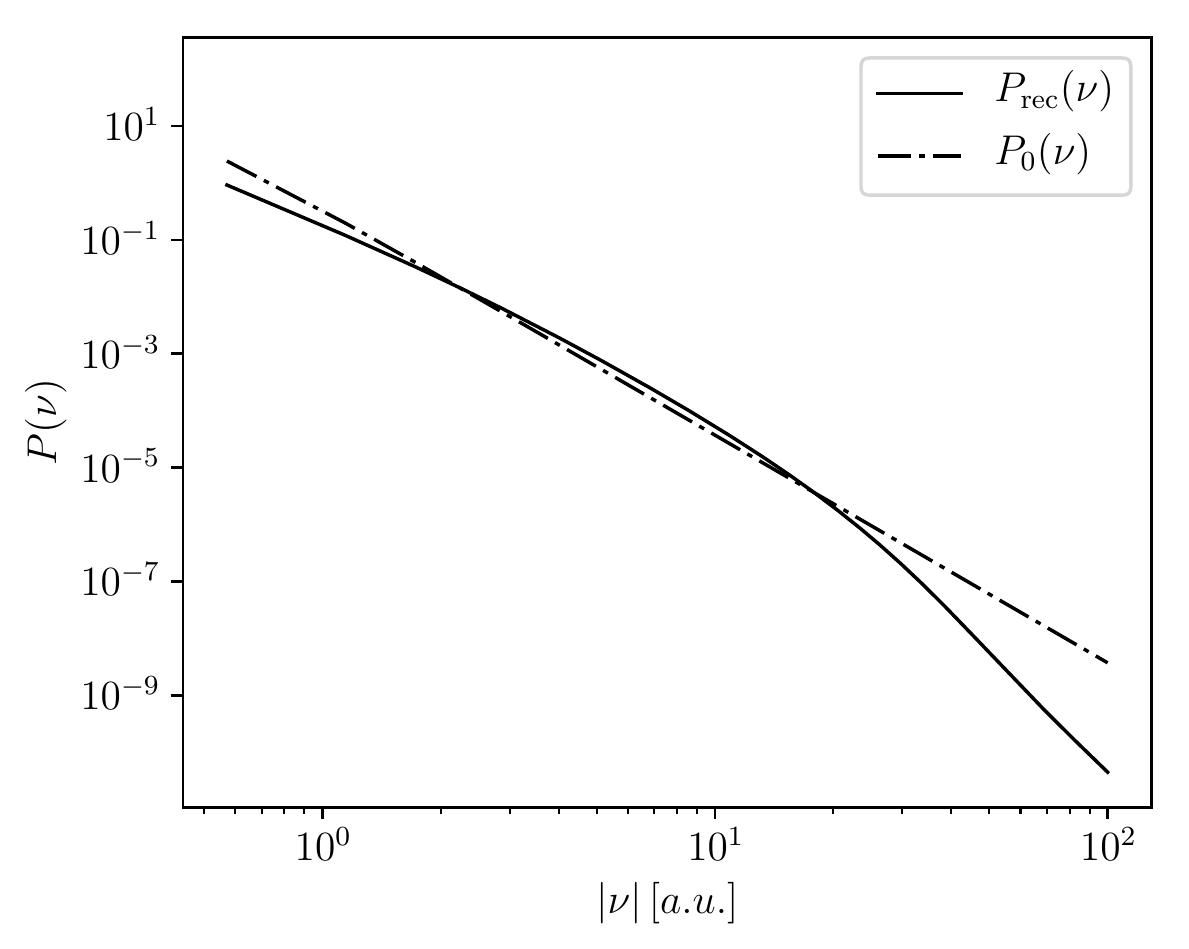}
		\caption{}
		\label{subfig:sub_s_0}
	\end{subfigure}
	\begin{subfigure}{6cm}
	\centering 
		\includegraphics[width=\textwidth]{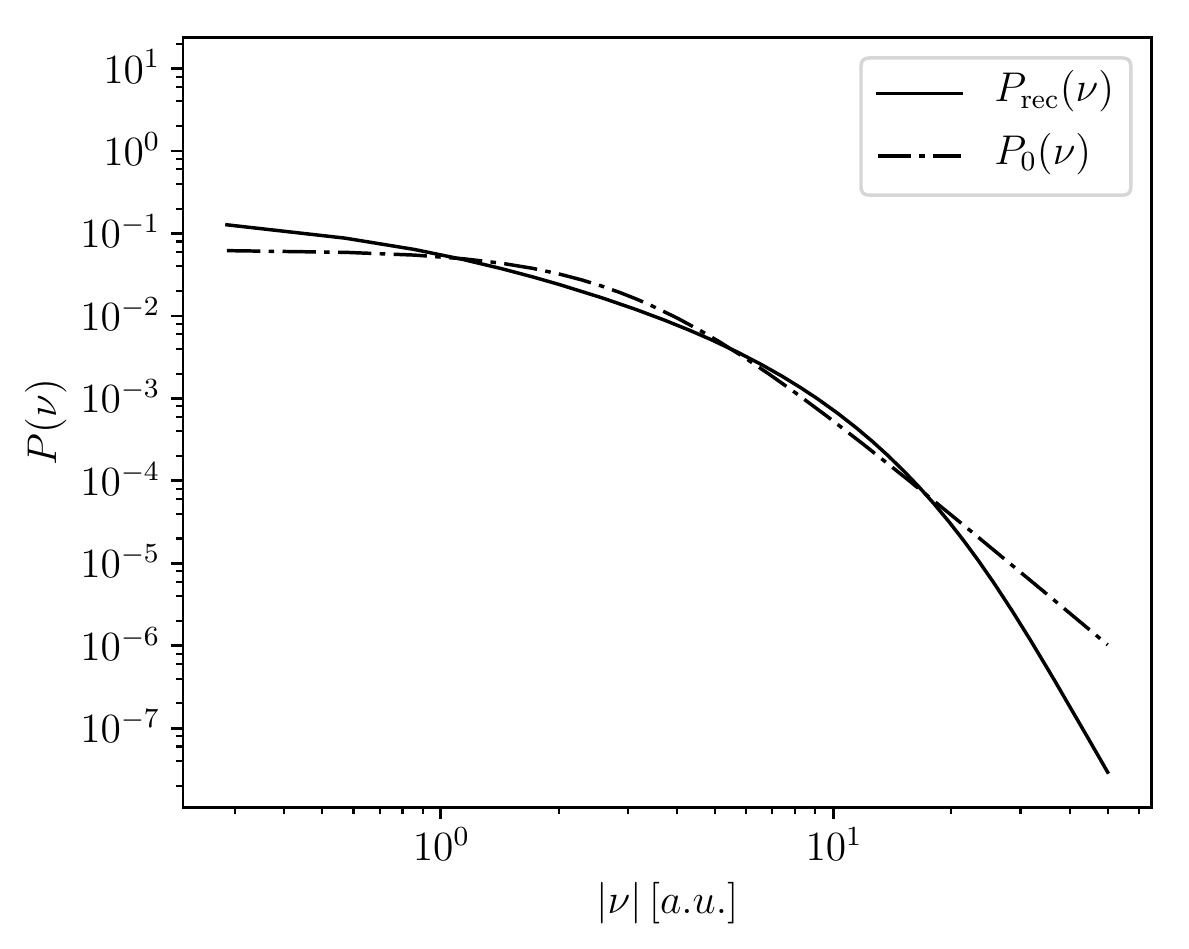}
		\caption{}
		\label{subfig:sub_s_1}
	\end{subfigure}	
	\begin{subfigure}{6cm}
	\centering 
		\includegraphics[width=\textwidth]{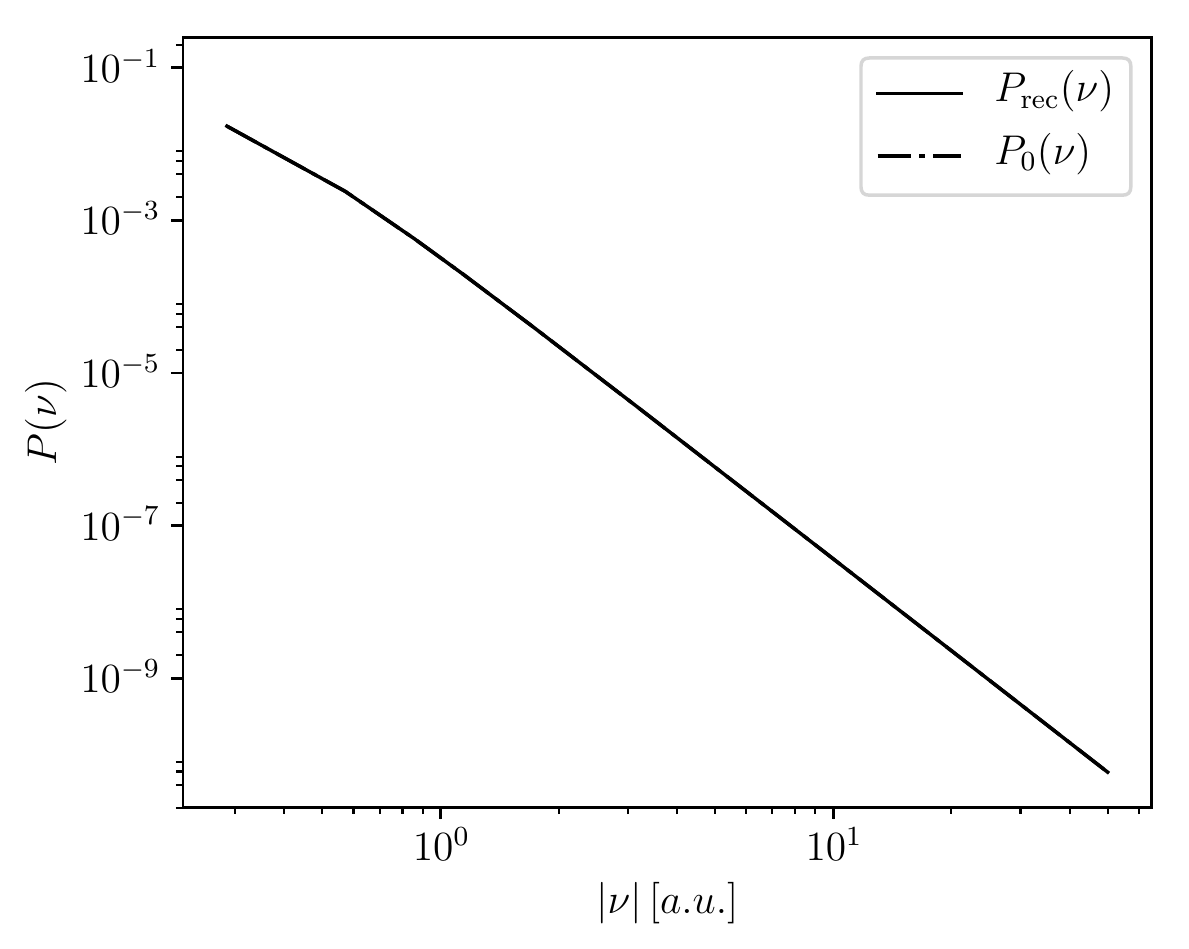}
		\caption{}
		\label{subfig:sub_u_1}
	\end{subfigure}

\caption{Illustration of the reconstructed power spectrum of $\vec{s}$ in its spatial (Fig.~\ref{subfig:sub_s_0}) and spectral sub-domain (Fig.~\ref{subfig:sub_s_1}) and $\vec{u}$ in its spectral domain (Fig.~\ref{subfig:sub_u_1}). The dashed black line indicates the default spectrum from which the Gaussian random fields, shown in Fig.~\ref{fig:sub_maps}, were drawn, while the solid black lines show its reconstruction. In case of $\vec{\tau}_{\mathcal{Y}^{\left(\vec{u}\right)}}$, both lines are in such close agreement that they are visually indistinguishable. }
\label{fig:specs}
\end{figure*}

From Fig.~\ref{fig:specs} it becomes apparent that the reconstructed power spectra of $\vec{s}$ track all large scale modes in good agreement up to $\nu \lesssim 20$. At higher harmonic modes the reconstructed power spectra start to deviate from the reference and fall more steeply. The drop off point at $\nu \approx 20$ roughly corresponds to the support of the PSF of the instrument's response. As the power spectrum still shows a smooth shape at $\nu \gtrsim 20$, the action of the smoothness enforcing prior starts to set in. Would $\sigma$ be significantly smaller than the used one, the spectra would start to scatter wildly, which we do not expect in astrophysical spectra. Hence the smoothness enforcing prior allows some kind of superresolution up to a certain threshold.\\
 
Having a closer look at the logarithmic point-like flux field (Fig.~\ref{fig:sub_maps}), we observe a similar situation as for diffuse flux field. This is supported by the reconstructed power spectrum, $\vec{\tau}_{\mathcal{Y}^{\left(\vec{u}\right)}}$ and its original one which match perfectly (Fig.~\ref{subfig:sub_u_1}). Up to where the reconstruction is mainly driven by the data may not read of as the algorithm recovered all modes correctly. This is of course due to an appropriate setup of the smoothness enforcing prior. \\

Nevertheless it must be noted that $\sigma$ (see Sec.~\ref{sec:Prior_smoothness})
 has to be set accurately as it can have significant influence on the reconstructed power spectrum. For a detailed discussion about its influence we refer to \cite{2017arXiv170805702P}.\\ 
\begin{figure}[]
\includegraphics[width=.48\textwidth]{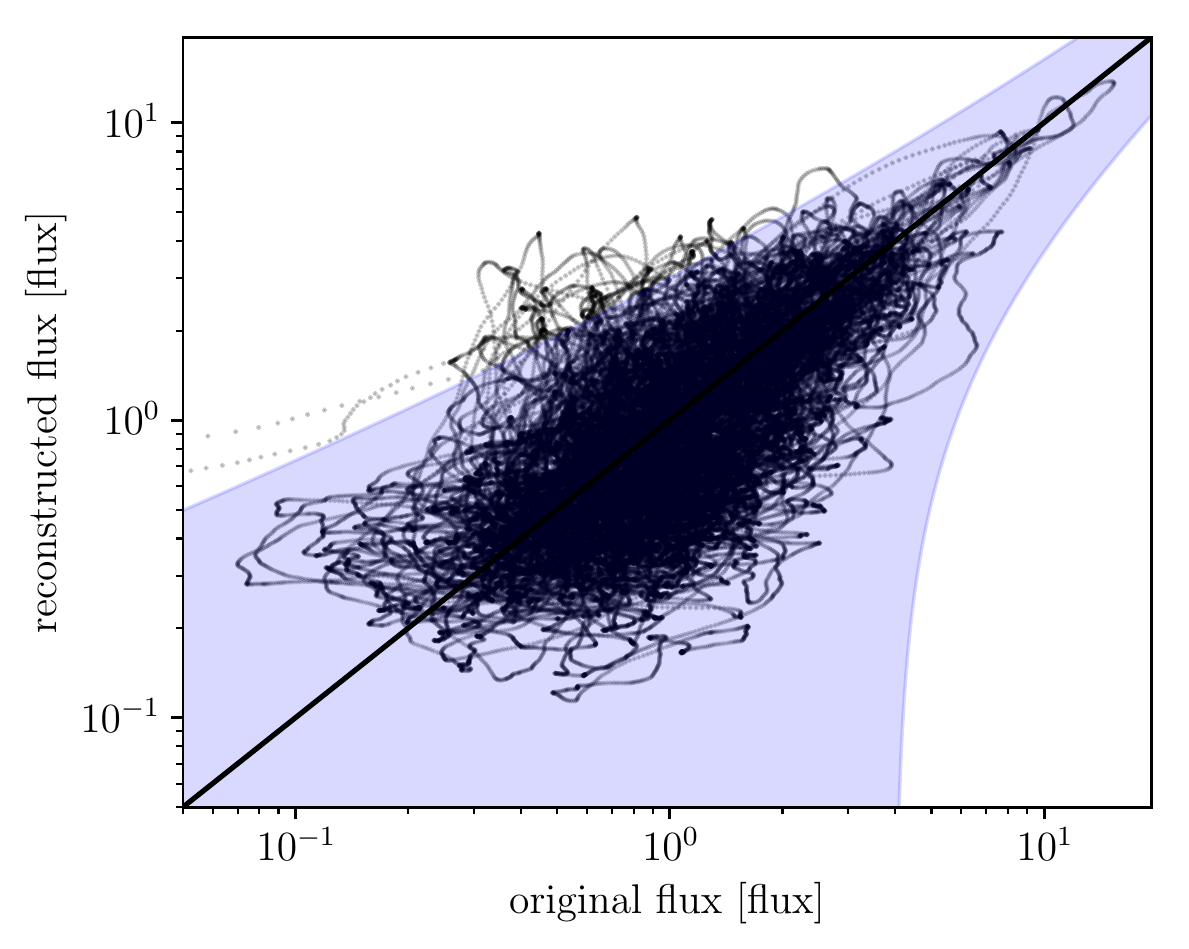}
\caption{Reconstructed versus original physical point-like flux in its spatial and spectral domain for all energies. The grey contour indicates the $2\sigma$ confidence interval of the Poissonian shot noise.}
\label{fig:scatter_sub}
\end{figure} 
The results for the spatial reconstruction performance of the point-like sources are plotted in Fig.~\ref{fig:scatter_sub}. For all energies we show individually the match between original and reconstructed flux at all locations. As a point-like source illuminates the sky over a broad range of energies at a fixed location these serpentine pattern appear in Fig.~\ref{fig:scatter_sub}. In total the reconstructed point-like flux is in good agreement within the $2\sigma$ confidence interval. This confidence interval corresponds to a diffuse and background free data set, it only illustrates the expected photon shot noise of point sources. The higher flux point-like sources tend to be reconstructed more accurately as a better S/N allows a sharper decomposition of the different sources. In the same way, the accuracy of the reconstruction for sources with low count rates becomes worse as the S/N becomes severe. The calculated absolute error supports these findings. \\
\begin{figure}[]
\includegraphics[width=.48\textwidth]{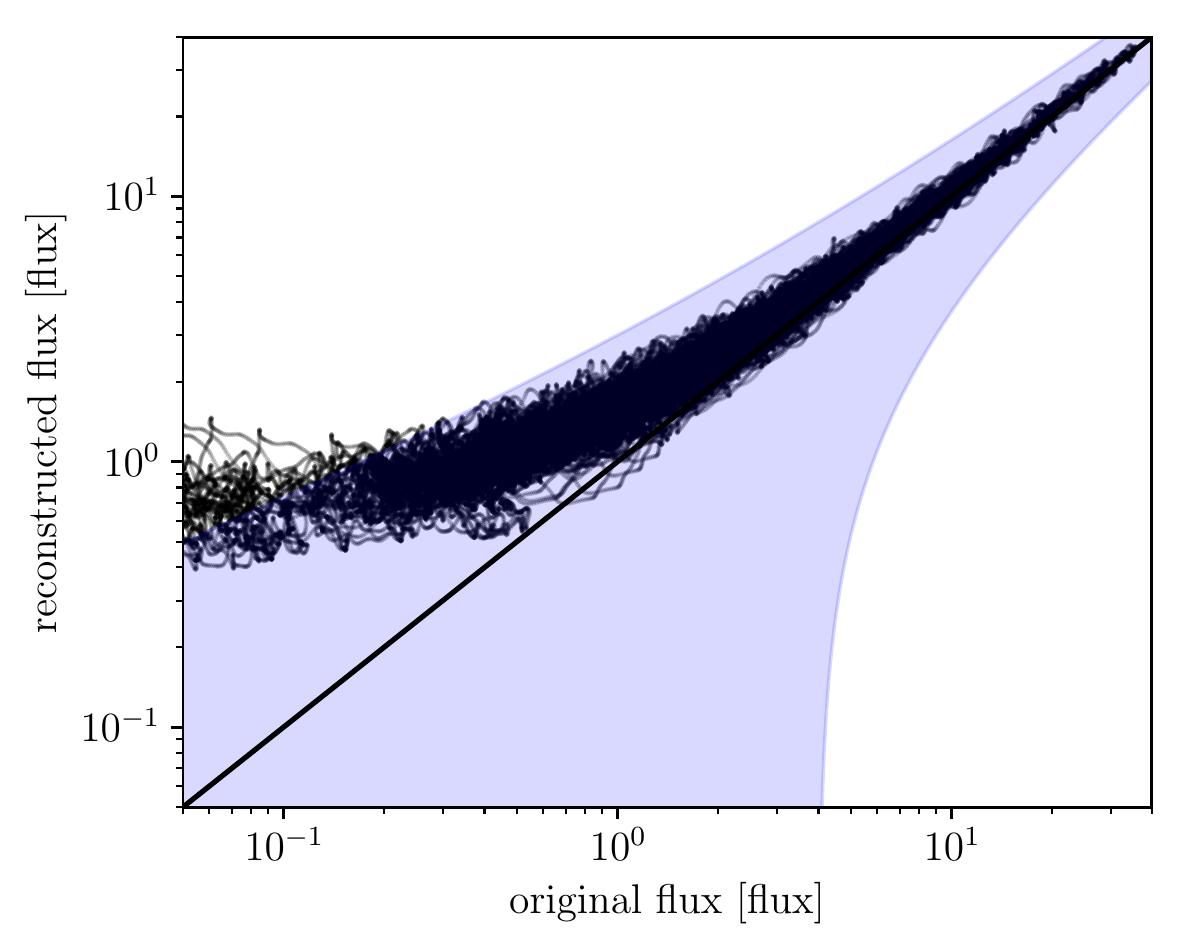}
\caption{Reconstructed versus original physical diffuse flux in its spatial and spectral domain for all energies. The grey contour indicates the $2\sigma$ confidence interval of the Poissonian shot noise.}
\label{fig:scatter_s}
\end{figure} 
Figure~\ref{fig:scatter_s} illustrates the quality of the spatial and spectral reconstruction of the diffuse flux. For all energies and locations we compare the reconstructed flux with its original one. The confidence interval corresponds, similar as in Fig.~\ref{fig:scatter_sub}, to a point-like and background free data set. Therefore it only shows the expected photon shot noise of diffuse flux and is consequently only a lower bound on the uncertainty. As already mentioned before, the reconstructed spectra (Fig.~\ref{fig:specs}) of the diffuse flux in both dimensions are not falling as steeply as the original one and therefore the reconstructed diffuse flux picks up small scale features from the background flux. Consequently the reconstructed flux is overestimated at low flux levels. This behaviour can also be observed in Fig.~\ref{fig:sub_maps} by looking at the error of the reconstructed diffuse flux, which tends to be larger in regions where the original diffuse flux was low. This indicates that small-scale features from the background radiation are misinterpreted as diffuse flux. This is an expected behaviour as the decomposition is primarily based on the correlation structure of each component. On the one hand, if these components show correlations on similar scale the decomposition becomes more and more degenerate. On the other hand, it becomes more a one-to-one correspondence if the scales of the correlations differ more significantly. It would be beneficial if they live over completely different domains as it actually is the case in many real world measurement situations. 
In total the results of the performed inference, demonstrated in Fig.~\ref{fig:sub_maps} display exactly this behaviour. The decomposition performs well on scales where the correlations of the components are different to each other in contrast to those where they are more similar.
As the correlation length of the background flux is similar to the support of the assumed PSF of the instrument's response, not all of its small features could be reconstructed even though its statistical properties, i.e. the power spectrum, were provided and not inferred from the data. Hence the reconstruction is smeared out as one would expect. Therefore the calculated absolute error is more finely grained and its absolute magnitude is smaller compared to the reconstruction. \\
In order to get a quantitative statement about the quality of the reconstructed fields we define a relative error $\epsilon$ for each component
\begin{align}
\epsilon^{\left( \mathrm{i}\right)} = \left \vert \vec{\rho} ^{\left( \mathrm{i}\right)}-\langle \vec{\rho}\rangle^{\left( \mathrm{i}\right)}\right \vert / \left \vert \vec{\rho}^{\left( \mathrm{i}\right)}\right \vert\,,
\label{eq:qa_measure}
\end{align}
with $i \in \{\vec{s}, \vec{u}, \vec{b}\}$. The errors are $\epsilon^{\left( \vec{s} \right)} \approx 2.910 \%, \epsilon^{\left( \vec{u} \right)} \approx 9.470 \%$, and $\epsilon^{\left( \vec{b} \right)} \approx 8.542 \%$. For a detailed discussion on how such measures can change the view on Bayesian inference algorithms we refer to \cite{2014InvPr..30k4004B}. \\

\begin{figure*}[]
\centering
\includegraphics[width=\textwidth]{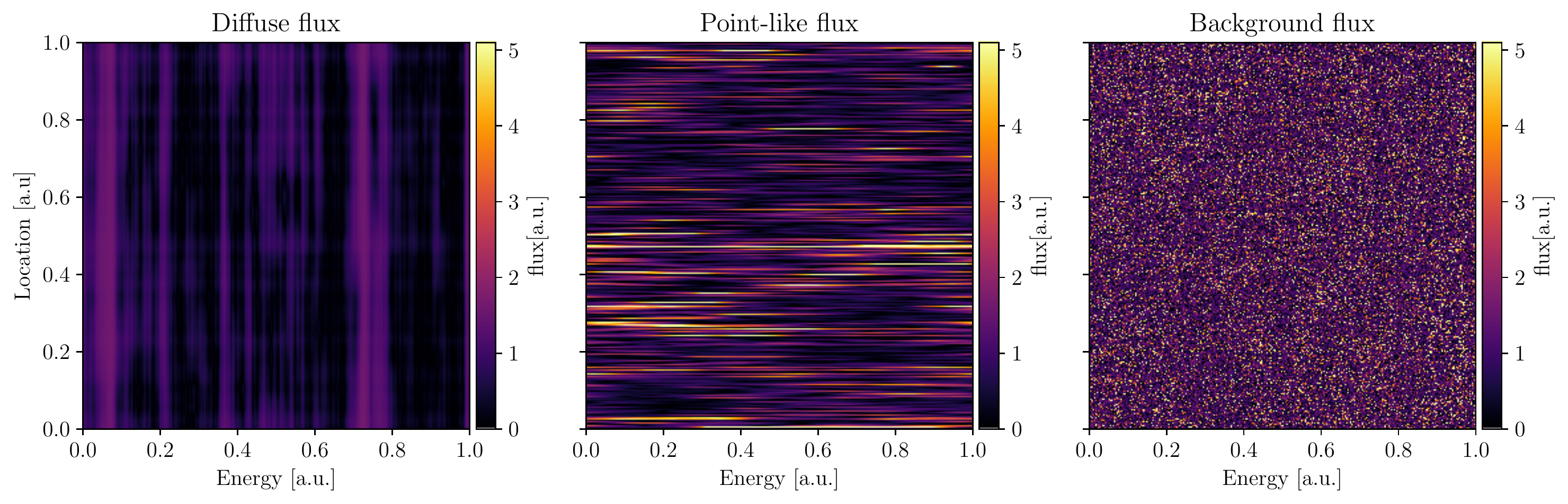}
\caption{The illustrations show the differences between $\left \vert \vec{\rho}^{\mathrm{error}} - \vec{\rho}^{\mathrm{uncertainty}}\right\vert$ on their native scale, by devision through the uncertainty estimates. The image-mean of the normalised errors is $\approx 0.6$ for the diffuse flux, $\approx 0.85$ for the point-like flux, and$\approx 1.5$  for the background flux. Hence they show, that some uncertainties are overestimated and others underestimated, respectively.}
\label{fig:uncert_use}
\end{figure*}

The calculated uncertainty estimates of the reconstructions, $\vec{s}$, $\vec{u}$, and $\vec{b}$ are as one would expect them to be, given the Poissonian shot noise and the Gaussian convolution of the response operator. The absolute magnitudes of the uncertainties are in all cases smaller than the amplitude of the reconstruction itself. Figure~\ref{fig:uncert_use} shows that uncertainty estimates are useful, however one has to keep in mind that approximation to the non-Gaussian posterior were done. Consequently for heavy tails of the posterior PDF the approximated uncertainty do not account for. \\

The inference algorithm is steered by the model parameters, shown in rectangular boxes in Fig.~\ref{fig:parameter_model}. In this mock data demonstration run we set them all to physically motivated values. Changing these, especially $\sigma$ of the smoothness enforcing prior and $\beta$, the shape parameter of the spatial prior for the point-like flux can have a significant influence on the reconstruction. A detailed parameter study of $\alpha, q, \beta$, and $\eta$ can be found in \cite{Selig:2015rt}. \\
It may be summarised that the advancements of D$^4$PO to denoise, deconvolve, and decompose multidimensional photon observations into multiple morphologically different sources, such as diffuse, point-like and background sources, while simultaneously learning their statistical properties in each of the field domains independently, have shown to give reliable estimates about the physical fluxes for data of sufficient quality. 

\subsection{Comparison to Maximum Likelihood methods (ML)}

\begin{figure*}[]
\centering
\includegraphics[width=\textwidth]{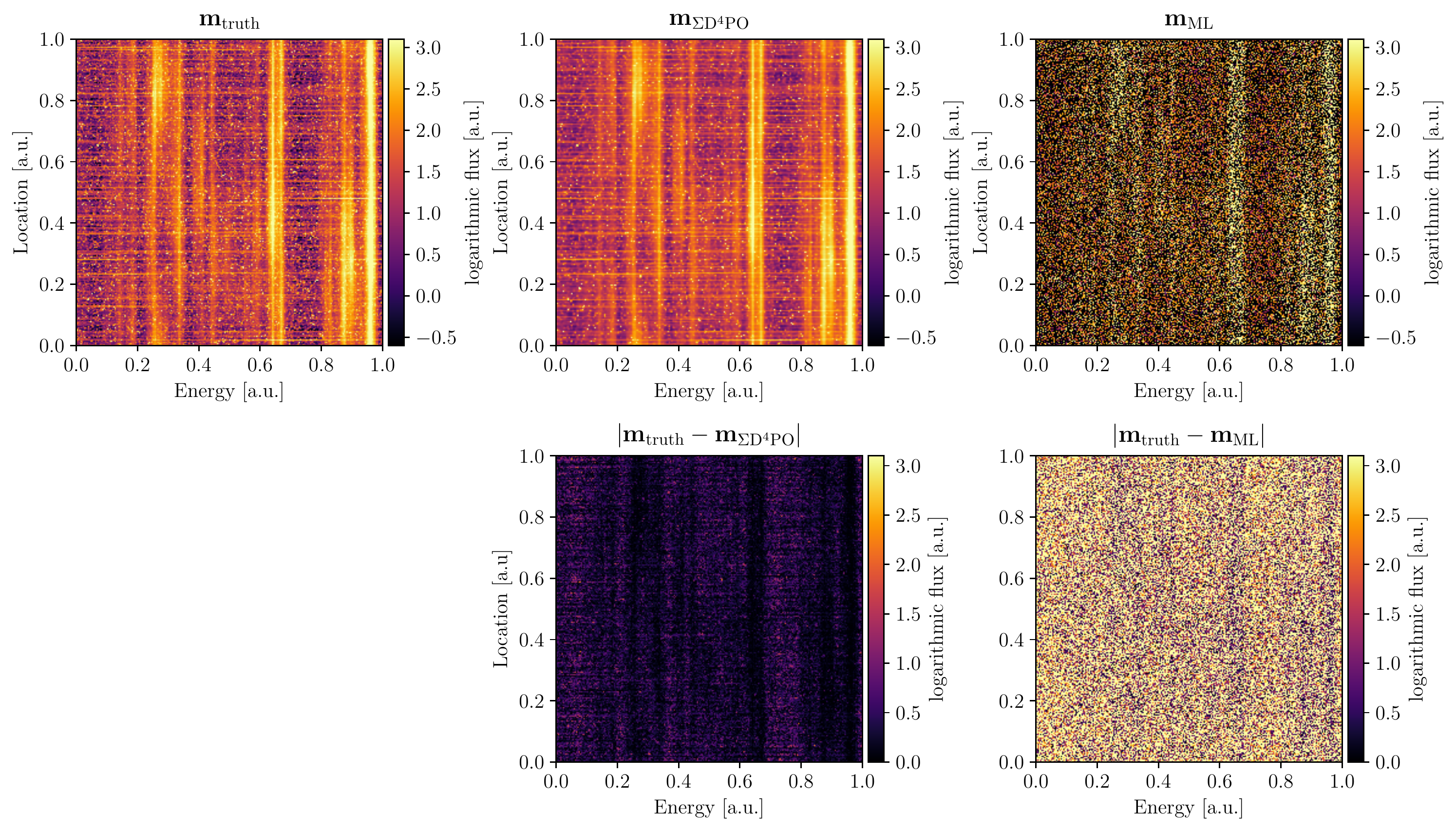}
\caption{The illustration indicates the differences between the introduced D$^4$PO reconstruction technique and a regular ML reconstruction. In the top left figure, the sum of all ground truth components (diffuse, point-like, and background flux fields), $\vec{m}_ \mathrm{ML}$ from Fig.~\ref{fig:sub_maps} is shown. The panel in the top middle shows the sum of the reconstructed flux components by D$^4$PO, $\vec{m}_{\Sigma \mathrm{ D^4PO}}$ (sum of reconstructions from Fig.~\ref{fig:sub_maps}), while the one at the right shows a regular maximum likelihood reconstruction: $\vec{m}_{\mathrm{ML}}$, using the same data set as in Fig.~\ref{fig:sub_maps}. Below are shown the absolute differences between the sum of all D$^4$PO reconstructions and the ground truth, $\vert \vec{m}_\mathrm{truth} -  \vec{m}_\mathrm{\Sigma\, D^4PO} \vert$ shown. The same applies for the maximum likelihood reconstruction,  $\vert \vec{m}_\mathrm{truth} -  \vec{m}_\mathrm{ML} \vert$ at the bottom right.}
\label{fig:comparison}
\end{figure*}

The maximum likelihood (ML) method is a well known technique to reconstruct fluxes in astrophysics \citep{Guillaume:98,2001,2001A&A...370..649B,1101146}. ML tries to purely maximise the likelihood (Sec.~\ref{subsec:likelihood}) and disregards any prior assumptions. Therefore minimising Eq.~(\ref{eq:Hamiltonian_likelihood}) is equal to maximising the likelihood and can only yield an estimate of the total photon flux of all components $\vec{\varphi}$. In Fig.~\ref{fig:comparison} exactly such a ML-reconstruction is shown and compared to a D$^4$PO reconstruction. To examine the differences between the two methods, we show the total sum of all components 
\begin{align}
\vec{m}_\mathrm{truth} = \log \left(e^{\vec{s}} +e^{\vec{u}} + e^{\vec{b}}\right)\,,
\end{align}
all reconstructed fluxes by D$^4$PO
\begin{align}
\vec{m}_\mathrm{\Sigma\, D^4PO} = \log \left(e^{\vec{s}_\mathrm{rec}} +e^{\vec{u}_\mathrm{rec}} + e^{\vec{b}_\mathrm{rec}}\right)\,,
\end{align}
(compare Fig.~\ref{fig:sub_maps}) and the reconstructed total flux by ML
\begin{align}
\vec{m}_{\mathrm{ML}} = \log (e^{\vec{\varphi}})\,.
\end{align}
From this comparison it becomes apparent that prior assumptions about the different 
photon flux components are not only essential to decompose a data set into morphologically different sources, but also for the quality of the reconstruction of the total flux. These break the degeneracy between different photon source components, especially when reconstructing and estimating more than one component. Further, Fig.~\ref{fig:comparison} shows that the reconstructed correlation structures of Fig.~\ref{fig:specs} help to recover more detailed features present in the original flux components ($\vec{s}, \vec{u}, \vec{b}$), than not using this information in a ML approach. Applying the same quantitative measure as in Eq.~(\ref{eq:qa_measure}), the relative error of $\vec{m}_\mathrm{\Sigma\, D^4PO}$ compared to $\vec{m}_\mathrm{truth}$ is $\epsilon^{\Sigma \mathrm{ D^4PO}} \approx 13.72 \%$, while the relative error of $\vec{m}_{\mathrm{ML}}$ is $\epsilon^{\mathrm{ML}} \approx 8896 \%$ . 

\section{Conclusion}
\label{sec:conclusion}
We derived the D$^4$PO algorithm, to denoise, deconvolve and decompose multi-domain photon observations, into multiple morphologically different components. In this context we focused on the decomposition of astrophysical high energy photon count data into three different types of sources, diffuse, point-like and background radiation. Each of these components lives over a continuous space over multiple domains, such as energy and location. In addition to the simultaneous reconstruction of all components the algorithm can infer the correlation structure of each field over each of its sub-domains. Thereby D$^4$PO takes accurate care of the instruments response function and the induced photon shot noise. Finally the algorithm can provide a posteriori uncertainty estimates of the reconstructed fields. \\

The introduced algorithm is based on D$^3$PO \citep{Selig:2015rt}, which was successfully applied to the FERMI LAT data \citep{Selig:2015ul} and giant magnetar flare observations \citep{2017arXiv170805702P}. D$^4$PO provides an advancement towards multidimensional fields analysis with multiple components. The D$^4$PO algorithm is based on the derived hierarchical Bayesian parameter model within the framework of IFT \citep{2009PhRvD..80j5005E}. The model incorporates multiple a priori assumptions for the signal fields of interest. These assumptions account for the known statistical properties of the fields in order to decompose the data set into their different sources. As some of these statistical identities are often not known a priori, the algorithm can learn them from the data set itself. Thereby we assume that each field follows a log-normal distribution over each of its sub-domains, except for the spatial correlations of point-sources. The correlations over these sub-domains may then be encoded via a power spectrum, implicitly assuming statistical homogeneity and isotropy over each sub-domain. As the point-like flux is implicitly defined to be statistically independent in its spatial sub-domain, we motivated an independent Inverse-Gamma prior, which implies a power law behaviour of the amplitudes of the flux. \\

To denoise and deconvolve astrophysical counts we took detailed care of an adequate likelihood modelling. The derived likelihood is a Poisson distribution to denoise the photon shot noise and incorporates all instruments artefacts, which are imprinted in the data set. \\

In total the hierarchical Bayesian parameter model is steered by only five parameters, for which we can provide well motivated a priori values. None of these parameters drives the inference dominantly as they may all be set to values where they provide minimal additional information to the inference problem. \\

In a simulated high energy photon count data set we demonstrated the performance of D$^4$PO. The algorithm successfully denoised, deconvolved, and decomposed the raw photon counts into diffuse, point-like and background radiation. Simultaneously it recovered the power spectrum of the diffuse flux in its spatial and spectral sub-domain. The correlation structure in the spectral sub-domain of the point-like flux was also inferred from the data. In total the analysis yielded detailed reconstructions and uncertainty estimates which are in good agreement with the simulated input. \\

The introduced algorithm is applicable to a wide range of inference problems. Its main advancement to reconstruct fields over multiple manifolds, each with different statistical identities, should find use in various data inference problems. Besides the most obvious applications in high energy astrophysics, such as FERMI, XMM-Newton, Chandra, Comptel, etc. to reconstruct the diffuse and point-like flux dependent on energy and location on the sphere, one may also infer energy dependent light curves of gamma ray bursts. Hence D$^4$PO has a broad range of applications.

\begin{acknowledgements}
We would like to thank Jakob Knollm{\"u}ller, Reimar Leike, Sebastian Hutschenreuter, Philipp Arras and an anonymous referee for their beneficial discussions during this project. 

\end{acknowledgements}
\bibliographystyle{aa}
\bibliography{bibliography}

\begin{appendix}
\section{Covariances and curvatures}
\label{app:curvature}
The covariance $D$ of a Gaussian $\mathcal{G} \left(\phi-\bar{\phi}, D\right)$ describes the uncertainty associated with the mean of the distribution. It may be computed via the inverse Hessian of the corresponding information Hamiltonian or Gibbs free energy respectively, 
\begin{align}
\left .\frac{\partial^2 G}{\partial \phi \partial \phi^\dagger}\right\vert_{\phi=\bar{\phi}} &= \left. \frac{\partial^2}{\partial \phi\partial \phi^\dagger} \left( \frac{1}{2}\left(\phi-\bar{\phi}\right )D^{-1} \left(\phi-\bar{\phi}\right)\right)\right\vert_{\phi= \bar{\phi}} \notag \\
& = D^{-1}\,.
\end{align}
The uncertainty covariances of the derived information Hamiltonian Eq.~(\ref{eq:Hamiltonian}) are, 
\begin{align}
D^{(\varphi)-1} &= \Phi_{\vec{t}}^{-1} + \left( 1- \frac{\vec{d}}{\vec{l}}\right)^\dagger \vec{\mathcal{R}}*\widehat{e^{\vec{m}}} +  \vec{\mathcal{R}}^\dagger \widehat{e^{\vec{m}}} \widehat{\frac{d}{l^{2}}}\vec{\mathcal{R}}^\dagger \widehat{e^{\vec{m}}}\,, \\
 D^{(\varphi_{\vec{u}_x})-1} &= D^{(\varphi)-1} + \vec{\eta} e^{-\vec{m}_{\vec{u}_x}} \,, \\
D^{(\vec{\tau})-1} &= \vec{T} + \left(q+\frac{w}{2}\right) e^{-\vec{t}}\,,
\label{eq:cur_H}
\end{align}
with 
\begin{align}
\vec{l} &= \vec{\mathcal{R}}e^{\vec{\varphi}} \,.
\end{align}
The corresponding covariances of the chosen Gibbs approach in Eq.~(\ref{eq:full_Gibbs}) are, 
\begin{align}
D^{-1} & = 
\begin{pmatrix}
D^{(\vec{\varphi})^{-1}} & 0  \\
 0 &  D^{(\vec{\tau})^{-1}}\\
 \end{pmatrix} \,, \quad \text{with} \\
D^{(\varphi)-1} &=\mathcal{T} \Bigg(  \Phi_{\vec{t}'}^{-1} + \left( 1- \frac{\vec{d}}{\vec{l}}\right)^\dagger \vec{\mathcal{R}}*\widehat{e^{\vec{m} + \frac{1}{2} \widehat{D}^{(\varphi)}}} \notag  \\ 
 &+ \vec{\mathcal{R}}^\dagger \widehat{e^{\vec{m}+ \frac{1}{2} \widehat{D}^{(\varphi)}}} \widehat{\frac{d}{l^{2}}}\vec{\mathcal{R}}^\dagger \widehat{e^{\vec{m}+ \frac{1}{2} \widehat{D}^{(\varphi)}}}\Bigg)\,, \\
 D^{(\varphi_{\vec{u}_x})-1} &= D^{(\varphi)-1} + \vec{\eta}\exp\left(-\vec{m}_{\vec{u}_x} + \frac{1}{2}\widehat{D}^{\left(\varphi_{\vec{u}_x}\right)}\right) \,, \\
D^{(\vec{\tau})-1} &= \mathcal{T}\left( \vec{T} + \left(q+\frac{w}{2}\right) e^{-\vec{t}'}\right)\,,
\label{eq:cur_Gibbs}
\end{align}
with 
\begin{align}
\vec{l} &= \vec{\mathcal{R}} e^{\vec{m}+ \frac{1}{2} \widehat{D}^{(\varphi)}}\,.
\end{align}
Up to the term $\frac{1}{2}\widehat{D}$ in the exponential functions, these covariances are identical to the ones derived via the maximum a posteriori ansatz Eq.~(\ref{eq:cur_H}). This shows that these higher order corrections terms change the uncertainty covariances, however their influence is hard to judge as they introduce terms that couple to all elements of $D$. 
It must be noted that the inverse Hessian only describes the curvature of the potential, hence it may only be regarded as the uncertainty of a reconstruction if the potential is quadratic. However, numerous numerical experiments showed that this assumption holds in most cases. 

\section{Deriving the Gibbs free energy}
\label{app:gibbs}
\subsection{The Gibbs free energy}
\label{subsec:Gibbs}
Due to the complex structure of the information Hamiltonian in Eq.~(\ref{eq:Hamiltonian}), we are seeking for an approximation to the true posterior Eq.~(\ref{eq:posterior_complete}). To this end we adapt a Gaussian distribution for our posterior approximation and require it to be close in an information theoretical sense to the correct posterior. The correct information distance is the Kullback-Leibler divergence, which is equivalent, up to irrelevant constants, to the Gibbs free energy \citep{2010PhRvE..82e1112E}. We directly adopt the final functional form of the posterior in the construction of the Gibbs free energy. Here we use an approximate Gaussian ansatz for the posterior Eq.~(\ref{eq:posterior_complete}) of our signal vector $\phi= \left(\vec{\varphi}^\dagger, \vec{\tau}^\dagger\right)^\dagger$:
\begin{align}
P(\phi \vert \vec{d}) &= \mathcal{G} \left(\phi-\bar{\phi}, D\right) \label{eq:free_gaussian} \quad \text{with}\\
\bar{\phi} &= \left( \vec{m}^\dagger, \vec{t}^\dagger\right)^\dagger \,,
\end{align}
the posterior mean, and 
\begin{align}
D=
\begin{pmatrix}
D^{(\vec{\varphi})} & 0 \\
0 &  D^{(\vec{\tau})} 
\end{pmatrix} 
\end{align}
the posterior uncertainty covariance. 
The posterior mean $\bar{\phi}$ consists of the mean field $\vec{m}= \langle \vec{\varphi}\rangle_{(\phi\vert d)}$, as well as the mean log power spectrum $\vec{t} = \langle \vec{\tau}\rangle_{(\phi\vert d)}$. 
The signal covariance $D = \langle(\phi-\bar{\phi}) (\phi-\bar{\phi})^\dagger\rangle_{(\phi\vert d)}$ consists of $2\times2$ block matrices, where the off diagonal terms are set to zero to reduce the complexity of the resulting algorithm. The non zero blocks are the signal uncertainty
\begin{align}
D^{(\vec{\varphi})}= \left \langle \left(\vec{\varphi} - \vec{m}\right)\left(\vec{\varphi} - \vec{m}\right)^\dagger\right\rangle_{(\phi\vert d)}\,, 
\end{align}
and the log-spectrum uncertainty
\begin{equation}
D^{\left( \vec{\tau}\right)}= \left \langle \left( \vec{\tau}- \vec{t}\right)\left( \vec{\tau}- \vec{t}\right)^\dagger \right \rangle_{(\phi \vert d)}\,.
\end{equation}
In terms of these parameters the Gibbs free energy is given by
\begin{equation}
G(\bar{\phi}, D \vert d)= U( \bar{\phi}, D \vert d) - \mathcal{T} \mathscr{S} (\bar{\phi}, D \vert d)\,,
\label{eq:def_Gibbs}
\end{equation}
with 
\begin{equation}
U(\bar{\phi}, D \vert d) = \left \langle H\left(\vec{\varphi}, \vec{\tau}\vert d \right)\right \rangle_{(\phi\vert d)}\,, 
\label{eq:internal_energy}
\end{equation}
being the internal energy, describing the full non-Gaussian Hamiltonian Eq.~(\ref{eq:Hamiltonian}), averaged by the approximated posterior Eq.~(\ref{eq:free_gaussian}). The entropy of the approximated Gaussian posterior is  
\begin{equation}
\mathscr{S}(\bar{\phi}, D \vert d)= - \int \mathcal{D}\phi\, P(\phi\vert d) \log P(\phi\vert d)\,. 
\label{eq:entropy}
\end{equation}
Up to an irrelevant sign and an additive constant the Gibbs free energy Eq.~(\ref{eq:def_Gibbs}) is equal to the Kullback-Leibler distance of the posterior approximation if $\mathcal{T}=1$. Unless stated differently, we therefore usually set $\mathcal{T}=1$. If $\mathcal{T}=0$, the approximate posterior given by Eq.~(\ref{eq:free_gaussian}) becomes a delta function at the maximum of the correct posterior, Eq.~(\ref{eq:posterior_complete}), which might perform poorly in case nuisance parameters, such as $\vec{\tau}$ also need to be reconstructed. 

In total the Gibbs free energy becomes

\begin{align}
G=& G \left ( \bar{\phi}, D \vert \vec{d} \right) \notag \\
=&  G_0 + \vec{1}^\dagger \vec{l} - \vec{d}^\dagger \log \vec{l} \notag\\
& + \frac{1}{2}  \Big[\Tr \left [ \log \Phi\right] +\vec{m}^\dagger\Phi^{-1} \vec{m} + \Tr\left(\Phi^{-1} D^{(\vec{\varphi})}\right) \Big]\notag \\
 & + \left( \alpha -1 \right)^\dagger \vec{t} + q^\dagger \exp \left(-\vec{t}+ \frac{1}{2}\widehat{D}^{(\vec{\tau})}\right) \notag \\
 & + \frac{1}{2} \vec{t}^\dagger \vec{T} \vec{t} + \frac{1}{2} \Tr \left[ \vec{T} D^{\left(\vec{\tau}\right)}\right] \notag \\
 & + \left(\beta-1\right)^\dagger \vec{m}_{\vec{u}_x} + \vec{\eta}^\dagger \exp\left(-\vec{m}_{\vec{u}_x} + \frac{1}{2}\widehat{D}^{\left(\varphi_{\vec{u}_x}\right)}\right)  \notag \\
& - \frac{\mathcal{T}}{2} \Tr \left [\mathds{1} + \ln \left( 2 \pi D\right) \right] \,,
\label{eq:full_Gibbs}
\end{align}
with $\Phi= \diag\left(\mathcal{S}, \mathcal{U}, \mathcal{B}\right)$, $G_0$ is absorbing all constants, and 
\begin{align}
\vec{l} &= \vec{\mathcal{R}} e^{\vec{m}+ \frac{1}{2} \widehat{D}^{(\varphi)}}\,.
\label{eq:l_gibbs}
\end{align}
Comparing Eq.~(\ref{eq:full_Gibbs}) with the information Hamiltonian Eq.~(\ref{eq:Hamiltonian}), there are a number of correction terms appearing which properly account for the uncertainty of the inferred map $\vec{m}$. In particular $\vec{l}$ differs, comparing Eq.~(\ref{eq:Hamiltonian_l}) with Eq.~(\ref{eq:l_gibbs}), which in the framework of Gibbs free energy describes the expectation value of $\vec{\lambda}$ over the approximate posterior Eq.~(\ref{eq:free_gaussian}). Minimizing the Gibbs free energy with respect to $\vec{m}, \vec{t}$ and $D$ would optimize the inference under the assumed Gaussian posterior. \\

In the following sections we will gradually derive the Gibbs free energy. 

\subsubsection{The Entropy}
Due to the Gaussian ansatz Eq.~(\ref{eq:free_gaussian}) the entropy Eq.~(\ref{eq:entropy}) is independent of $\bar{\phi}$,
\begin{equation}
\mathscr{S}(\bar{\phi}, D\vert d)= \frac{1}{2} \Tr \left [ \mathds{1} + \ln (2\pi D)\right] 
\label{eq:entropy_calculated}
\end{equation}
and therefore its gradients with respect to $\vec{m}$ and $\vec{t}$ vanish, 
\begin{align}
\frac{\partial \mathscr{S}}{\partial \vec{m}}= 0\,, \quad 
\frac{\partial \mathscr{S}}{\partial \vec{t}}= 0\,.
\end{align}

\subsubsection{Internal energy of the hyperprior}

The internal energy of the hierarchical Gaussian prior $P(\vec{\tau} \vert \sigma, \alpha, q)$ is
\begin{align}
U^{\left(\vec{\tau}\right)}\left( \bar{\phi}, D \vert \vec{d}\right) =& \left \langle H(\vec{\tau})\right \rangle_{(\phi, D\vert \vec{d})} \\
\simeq& \frac{1}{2} \vec{t}^\dagger \vec{T} \vec{t} + \frac{1}{2} \Tr \left[ \vec{T} D^{\left(\vec{\tau}\right)}\right] \notag \\
& + \left(\alpha -1\right)^\dagger \vec{t} + q^\dagger e^{-\vec{t} + \frac{1}{2}\widetilde{D}^{\left(\vec{t}\right)}}\,,
\label{eq:Hyperprior_energy}
\end{align}
with $\vec{T} = (T_{\mathcal{Y}^{(s)}}, T_{\mathcal{X}^{(s)}}, T_{\mathcal{Y}^{(u)}}, T_{\mathcal{Y}^{(b)}}, T_{\mathcal{I}^{(b)}})^\dagger$. A hat on a tensor denotes the diagonal vector in the position basis, $\widehat{D}_x= D_{xx}$, while a hat on a vector refers to a tensor with the vector as its diagonal, $\widehat{m}_{xy} = \delta_{xy} m_x$. Similarly we define a tilde on a tensor as the diagonal vector in the band harmonic basis, $\widetilde{\theta}_k= \theta_{kk}$, and a tilde on a vector denotes a tensor with the vector on its diagonal in the band harmonic basis, $\widetilde{t}_{kl} = \delta_{kl} t_k$.  

The corresponding non-vanishing gradients of the hyperprior's internal energy Eq.~(\ref{eq:Hyperprior_energy})
are 
\begin{align}
\frac{\partial U^{\left(\vec{\tau}\right)}}{\partial \vec{t}} =&   \vec{T} \vec{t} + \left( \alpha -1\right) - q e^{-\vec{t}'}\,, \label{eq:U_int_t}\\
\frac{\partial U^{\left(\vec{\tau}\right)}}{\partial D^{\left(\vec{\tau}\right)}}=& \frac{1}{2} \left [ \vec{T} + \widetilde{q e^ {-\vec{t}'}}\right]\,, \label{eq:U_int_Dt}
\end{align}
with $\vec{t}'= t- \frac{1}{2}\widetilde{D}^{(\vec{\tau})}$.

\subsubsection{Internal energy of the prior for $\vec{\varphi}$}
The priors for $\vec{\varphi}$ provide the internal energy
\begin{align}
U^{\left(\vec{\varphi}\right)}(\bar{\phi}, D\vert \vec{d}) &=\, \frac{1}{2}\Tr \left [ \log \Phi_{\vec{t}'}\right] \notag \\
&+  \frac{1}{2}\left[ \vec{m}^\dagger\Phi_{\vec{t}'}^{-1} \vec{m} + \Tr\left[\Phi_{\vec{t}'}^{-1} D^{(\vec{\varphi})}\right]\right]\,.
\end{align}
The corresponding gradients are
\begin{align}
\frac{\partial U^{\left(\vec{\varphi}\right)}}{\partial \vec{m}} =& \Phi_{\vec{t}'}^{-1} \vec{m} \,, \label{eq:U_p_m}\\
\frac{\partial U^{\left(\vec{\varphi}\right)}}{\partial D^{({\vec{\varphi})}}} =& \frac{1}{2} \Phi_{\vec{t}'}^{-1} \, \label{eq:U_p_D}\\
\frac{\partial U^{\left(\vec{\varphi}\right)}}{\partial t_{\mathcal{Y}}} =& \frac{1}{2} \left ( \rho_{\mathbb{P}} -  w_{t'_{\mathcal{Y}}}e^{-t'_{\mathcal{Y}}} \right)\,, \label{eq:U_p_ty}\\
\frac{\partial U^{\left(\vec{\varphi}\right)}}{\partial t_{\mathcal{X}}} =& \frac{1}{2} \left ( \rho_{\mathbb{P}} - w_{t'_{\mathcal{X}}}e^{-t'_{\mathcal{X}}} \right) \,,\label{eq:U_p_tx} \\
\frac{\partial U^{\left(\vec{\varphi}\right)}}{\partial D^{ \left(t_{\mathcal{X}}\right)}} =& \frac{1}{4} \left( \widetilde{w_{t'_{\mathcal{X}}}e^{-t'_{\mathcal{X}}} }\right)\,,\label{eq:U_p_Dx} \\
\frac{\partial U^{\left(\vec{\varphi}\right)}}{\partial D^{\left(t_{\mathcal{Y}}\right)}} =& \frac{1}{4} \left ( \widetilde{w_{t'_{\mathcal{Y}}}  e^{-t'_{\mathcal{Y}}}} \right) \,, \label{eq:U_p_Dy}
\end{align} 
with 
\begin{align}
w_{t'_{\mathcal{Y}}}&= \Tr \left[ \mathcal{X}^{(\vec{\varphi})^{-1}} \mathbb{P}\left( \vec{m}^\dagger \vec{m} + D^{(\vec{\varphi})}\right)\right] \,,\\
w_{t'_{\mathcal{X}}}&= \Tr \left[ \mathcal{Y}^{(\vec{\varphi})^{-1}} \mathbb{P}\left( \vec{m}^\dagger \vec{m} + D^{(\vec{\varphi})}\right)\right] 
\end{align}
and with the multiplicities $\rho_{\mathbb{P}}= \Tr \mathbb{P}$ of the spectral bands $k$ and $l$, which are sharing the same harmonic eigenvalue.

\subsubsection{Internal energy of the Inverse-Gamma prior}
The internal energy of the inverse-gamma prior on $\vec{\varphi}_{\vec{u}_x}$ is 
\begin{align}
U^{\left(\vec{\varphi}_{\vec{u}_x}\right)}(\bar{\phi}, D\vert \vec{d}) &= \left(\beta-1\right)^\dagger \vec{m}_{\vec{u}_x} + \vec{\eta}^\dagger \exp\left(-\vec{m}_{\vec{u}_x} + \frac{1}{2}\widehat{D}^{\left(\varphi_{\vec{u}_x}\right)}\right) \,,
\label{eq:U_ig}
\end{align}
with its corresponding gradient
\begin{align}
\frac{\partial U^{\left(\vec{\varphi}_{\vec{u}_x}\right)}}{\partial \vec{m}_x} &= \vec{\beta} - 1 -\vec{\eta} *  \exp\left(-\vec{m}_{\vec{u}_x} + \frac{1}{2}\widehat{D}^{\left(\varphi_{\vec{u}_x}\right)}\right) \,.
\label{eq:U+ig_d_m}
\end{align}

\subsubsection{Internal energy of the likelihood}
The internal energy of the likelihood Eq.~(\ref{eq:Hamiltonian_likelihood}) is
\begin{align}
U (\bar{\phi}&, D \vert \vec{d}) = \left \langle H(d \vert \vec{\varphi}, D)\right \rangle_{(\phi, D \vert \vec{d})} \notag \\ 
\simeq&\, G_0 + \vec{1}^\dagger \vec{l} \notag \\
&\,-\vec{d}^\dagger \left \{ \log (\vec{l}) - \sum_{\nu=2}^\infty \frac{(-1)^\nu}{\nu} \left \langle \left( \frac{\vec{\lambda}}{\vec{l}} - 1\right)^\nu \right \rangle \right\}\,,
\label{eq:Gibbs_likelihood}
\end{align} 
with
\begin{align}
\vec{\lambda}&= \vec{\mathcal{R}}e^{\vec{\varphi}}   \,, \label{eq:Gibbs_lambda}\\
\vec{l} &= \vec{\mathcal{R}} e^{\vec{m}+ \frac{1}{2} \widehat{D}^{(\varphi)}} \,, \label{eq:Gibbs_l}
\end{align}
where we absorbed all terms that are constant in $\vec{\varphi}$ into $G_0$. The evolution of the internal energy would require to know all entries of $D^{\left(\varphi\right)}$ explicitly. As this is computationally infeasible and the term  $\left \langle \left( \frac{\vec{\lambda}}{\vec{l}} - 1\right)^\nu \right \rangle \approx 0$ at the mode where $\lambda \approx l$, this sum can be neglected. As a consequence all cross correlations, such as $D^{\left(\vec{su}\right)}$ are implicitly set to zero. \\
The gradient of Eq.~(\ref{eq:Gibbs_likelihood}) can be derived by taking the first derivative of $U (\bar{\phi}, D \vert \vec{d})$ with respect to the approximate mean estimates, 
\begin{align}
\frac{\partial U}{\partial \vec{m}}  =& \left( 1- \frac{\vec{d}}{\vec{l}}\right)^\dagger \vec{\mathcal{R}}* e^{\vec{m} + \frac{1}{2} \widehat{D}^{(\varphi)}}  \, . \label{eq:G_ms_grad}
\end{align} 

\end{appendix}

\end{document}